\documentstyle[prl,aps,epsfig]{revtex}

\begin{document}
\def\be{\begin{equation}}
\def\ee{\end{equation}}
\def\bea{\begin{eqnarray}}
\def\eea{\end{eqnarray}}
\def\bml{\begin{mathletters}}
\def\eml{\end{mathletters}}
\def\l{\label}
\def\eqn#1{(~\ref{eq:#1}~)}
\def\no{\nonumber}
\def\av#1{{\langle  #1 \rangle}}

\title{Phases of a conserved mass model of aggregation with
fragmentation at fixed sites} 

\author{Kavita Jain and Mustansir Barma}
\address{Department of Theoretical Physics, Tata
Institute of Fundamental Research\\
Homi Bhabha Road, Mumbai 400005, India.}
\maketitle
\widetext

\begin{abstract}
To study the effect of quenched disorder in a class of
reaction-diffusion systems, we introduce a conserved mass model of
diffusion and 
aggregation in which fragmentation occurs only at certain fixed
sites. On most sites, the mass moves as a whole to a
nearest neighbour while it leaves the fixed sites only as a single
monomer (i.e. chips off). Once the mass leaves any site, it coalesces
with the mass present on its neighbour. The total mass of the system
is conserved in this model as is evident from the rules.
We study in detail the effect of a \emph{single} chipping site on the 
steady state in arbitrary dimensions, with and without bias.
In the thermodynamic limit, the system can exist in one of the
following three phases --
(a) Pinned Aggregate (PA) phase in which an infinite aggregate
(i.e. an aggregate with mass proportional
to the volume of the system) appears at the chipping site with 
probability one but not in the bulk. (b) Unpinned Aggregate (UA) phase
in which $\emph{both}$ the chipping site and the bulk can support an 
infinite aggregate simultaneously.
(c) Non Aggregate (NA) phase in which the system 
cannot sustain a cluster of mass proportional to the volume at all.
The steady state of the system depends on the dimension and
drive. A sitewise inhomogeneous Mean Field Theory predicts that the
system exists in the UA phase in all cases. We tested this prediction
by Monte Carlo simulations in $1d$ and $2d$ and found that it is true
except in $1d$, biased case.
In the latter case, there is a
phase transition from the NA phase to the PA phase as the mass density
is increased. We identify the critical point exactly and calculate the mass
distribution in the PA phase. The NA phase and the
critical point are studied by Monte-Carlo simulations and using
scaling arguments. A variant of the above aggregation model is also
considered in which total particle number is conserved and
chipping occurs at a fixed site, but the particles do not interact
with each other at
other sites. This model is solved exactly by mapping it to a Zero
Range Process. With increasing density, it exhibits a phase transition
from the NA phase to the PA phase in all dimensions, irrespective of
bias. The model is also solved with an extensive number of chipping
sites with random chipping rates and we argue that the solution
describes qualitatively the behaviour of the aggregation model with
extensive disorder.

\end{abstract}

\section{Introduction}

Reaction-Diffusion systems form an important class of nonequilibrium
systems whose dynamics and steady state
depend on various factors such as the nature of the reaction (aggregation,
annihilation, birth, fragmentation), number or
type of reactants involved (single or multi-species), velocity of
reactants (ballistic or diffusion controlled) and presence of external
input (injection) \cite{ziff,fried,priv} .
An interesting class with wide ranging applications involves the
elementary moves of $\emph{aggregation}$ (coalescence on contact) and
$\emph{fragmentation}$ (break-up of clusters of masses), besides
diffusion. A number of analytical results
including the occurrence of nonequilibrium phase transitions have been
obtained for such systems with translationally invariant geometries 
\cite{advec,drywet,inout,cmam,pbdry}.
A natural question arises -- what is the effect of quenched disorder on
the possible phases of such systems and the transitions between them? We
may anticipate interesting effects, as quenched disorder
is known to strongly influence the character of the
steady state in other nonequilibrium systems
\cite{kf,bec,gt,pwdis,krugR,evansR}. For  
instance, the steady state of driven diffusive systems on a
one-dimensional lattice with bias shows phase separation
in the presence of even a single defect \cite{bec,jl,schutz,mallick}.
 
To see what effect quenched disorder might have in a system with
diffusion and aggregation, consider the process of polymerisation in a
random medium with traps at certain fixed sites in which the polymer
can get stuck. Aggregation occurs when two diffusing chains meet and
coalesce; the reduced mobility of the aggregates at the trap sites promotes the
formation of large, \emph{localised} aggregates at such sites. If these traps
are not perfect and allow monomers to detach and leave, there is also
a possibility of formation of large \emph{mobile} aggregates in the bulk. 

To elucidate under what circumstances which types of
aggregates may form as a result of these physical effects, namely,
diffusion-aggregation, chipping and trapping, we consider a simple,
reduced model in which the mass leaves as a single aggregate to a
nearest neighbour from all sites except at certain, fixed sites from
which it is allowed to leave only by single monomer dissociation. 
We refer to the loss of unit mass as chipping and
these special sites as chippers. The total mass in the system is
conserved as is evident from the dynamical rules described above.
The quenched character of the chipper sites is important in
determining the
steady state of this model which is quite different from that in the
translationally invariant uniform-chipping model where chipping can
occur at every site \cite{cmam}.

As a first step towards understanding such a spatially inhomogeneous
system, we study in detail the case when only a $\emph{single}$
chipper is present. Despite the simplicity of the model, we find that
that the presence of a chipper gives rise to interesting steady
states. We study the model both analytically and numerically in arbitrary
dimensions, both with and without a global bias which sets up an overall
mass current. The constraint of conservation of total mass in the model
plays an important role in determining the steady state. We
find that the system may exist in three possible phases. Depending on the
dimension and the presence of drive, the system either stays in one of
the phases or else 
make transitions from one phase to the another as the parameters are
varied. These phases are 
characterised by the presence or absence of aggregates with mass
proportional to the volume of the system. In the thermodynamic limit, the
mass of such an aggregate diverges, so we refer
to it as an infinite aggregate. The three phases are described below:

\textbf{Pinned Aggregate phase (PA phase)}:
In this phase, with probability one, an infinite aggregate occurs only at
the chipper site and not in the bulk. This infinite aggregate acts as a
spatially localised particle bath for rest of the system.

\textbf{Unpinned Aggregate phase (UA phase)}: In this phase, a localised
infinite aggregate can exist at the chipper site together with mobile
infinite aggregates in the bulk. Interestingly, $\emph{both}$ types of
infinite aggregate can exist simultaneously. In this sense, this
state is different from  the aggregate phases in translationally
invariant systems such as the
no-chipper limit of this model or the uniform chipping model\cite{cmam},
in which only one mobile infinite aggregate occurs at a time.

\textbf{Non Aggregate phase (NA phase)}: This phase is characterised by
the absence of an infinite aggregate anywhere in the system. The mass is
spread out all over the system in clusters, each of which has a
vanishing fraction of the total mass in the thermodynamic limit.

We analysed the system within a Mean Field Theory (MFT) allowing for the
spatial dependence in the mass distributions. We find that the MFT
predicts the existence of only the UA phase. On comparing this result
with Monte Carlo simulations in $d=1$ and $d=2$, we find qualitative
agreement except in the $1d$, biased case. In this exceptional case,
there is a phase 
transition from the NA phase to the PA phase as the density is increased.
This case is different from the rest due to an
interesting interplay of the ballistic scale of motion and the diffusive
scale of coalescence in one dimension.

In the model described above, an infinite aggregate is formed due to both
interactions  (coalescence of diffusing particles) and the presence of an
inhomogeneity in the form of the chipper. It is useful to contrast the
behaviour of this model with that of a
model of $\emph{noninteracting}$ particles in which the aggregate is
formed solely due to the presence of disorder.
We solve this model in the presence of a single defect by mapping it
to a Zero Range 
Process \cite{evansR,spitzer} and show that it exhibits a phase
transition from the NA phase to the PA phase in all dimensions,
irrespective of bias. This is in contrast to the aggregation model
described earlier
which does not show a phase transition in higher dimensions in the
presence of single chipper.

Although we mainly discuss the case of a single chipper in this paper,
we also discuss what happens in the presence of extensive disorder in
both the models. We argue that in the presence of an extensive number
of chippers, 
the interaction effects (i.e. coalescence) can be ignored in the
aggregation model on large enough length and time scales and it behaves
like the free-particle model. The 
latter model can be solved in the presence of extensive disorder and
shows a phase transition in all dimensions for all bias from the PA
phase to the NA phase as the density is decreased.

The remainder of the paper is organised as follows. We define the
single-chipper aggregation model in Section~\ref{aggr} and discuss the
possible phases on
the basis of the conservation law. We analyse the system within a MFT in
Section~\ref{mft} and show that it predicts the occurrence of the UA
phase. We also discuss the numerical results which support this broad
conclusion in several cases. In Section~\ref{except} we turn to the
exceptional case namely the one dimensional, biased case. We present
analytical results in the PA phase and numerical results in the NA phase and
at the critical point. In Section~\ref{free}, we present the solution of the
single-chipper free particle model. We also discuss the likely
behaviour of the system with extensive disorder
in Section~\ref{extens}. Finally we summarise our
results in Section~\ref{concl}.

\section{Single-Chipper Aggregation model and its phases}
\label{aggr}

\subsection{The Model}
\label{dam}
Our model is defined on a $d$-dimensional hypercubic lattice of length
$L$ with periodic boundary conditions, and is studied both in
the presence and absence of bias. The system evolves via following rules:
at any site except the origin, 
\begin{figure}
\begin{center}
\psfig{figure=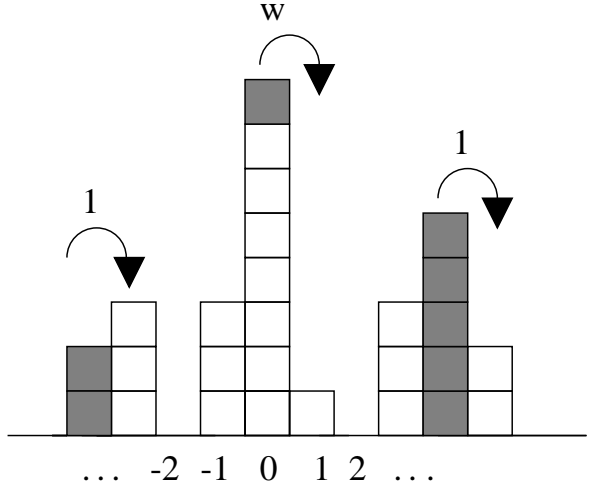,width=10cm,angle=0}
\caption{Illustration of the aggregation and chipping moves in
the totally biased case in one dimension. In an elementary move, only
a monomer (shaded) can leave the chipper site (located at the origin) at a
rate $w$. At sites other than the origin, the mass leaves as a whole (shaded)
at a rate $1$. Once the mass moves from any site to its neighbouring
site, it coalesces with the mass already present on that site.}
\label{model}
\end{center}
\end{figure}
\noindent all the mass at that site moves as a
whole to a nearest neighbour 
at a rate $1$, while at the origin only a single monomer chips off at
a rate $w$ leaving rest of the mass behind. Once 
the mass moves from any site to its neighbouring site, it simply coalesces
with the mass present on that site (see
Fig.~\ref{model}).
Evidently, the total mass  $M = \rho L^d$ of the system
is conserved, where $\rho$ is the mass density.  
The evolution of the system can be described
by the time evolution equation for the probability $P_{k}(m,t)$ that
there is mass $m$ at site $k$ at time $t$. As the moves at the origin
are different from 
that at the rest of the system, the mass distribution at the origin (denoted by
$0$) and its 
nearest neighbours which receive mass from the origin (denoted by set
$A$) obey equations different from those obeyed by rest of the system
(set $B$).

The set of time evolution equations obeyed by the sites in the bulk (set $B$)
are given by  
\bml
\bea
\frac{\partial P_k(m,t)}{\partial t} & = &\sum_{\delta} [ \sum_{m^\prime=1}^{m}
P_{k,k+\delta}(m-m^\prime,m^\prime,t) - \sum_{m^\prime \neq 0}
P_{k,k+\delta}(m,m^\prime,t) - P_{k}(m,t) ]  \;\;,\;\;k \in B \l{e11}  \\  
\frac{\partial P_{k}(0,t)}{\partial t} & = & \sum_{\delta}[\sum_{m
\neq 0} P_{k}(m,t)- 
\sum_{m^\prime \neq 0} P_{k,k+\delta}(0,m^\prime,t) ] 
\;\;,\;\; k \in B \;\;\;\;, \l{e12}
\eea
\eml
where $P_{k,k+\delta}(m,m^\prime,t)$ is the joint probability that site
$k$ and its nearest neighbour $k+\delta$ have mass $m$ and
$m^\prime$ respectively. The index $\delta$ runs over the 
$2d$ nearest neighbours in the unbiased case and $d$ forward
neighbours for totally biased case. 
The convolution term in Eq.(\ref{e11}) is the gain
term through which the deficit mass is supplied to site $k$ via
diffusion-aggregation moves from its nearest neighbours.
The gain term for zero mass in Eq.(\ref{e12}) results from hopping the mass
from site $k$ to one of its nearest neighbours. The system can get out of
the configuration in which the site $k$ has mass $m$ (including zero) 
when (a) the mass hops into site $k$ from its nearest neighbours and 
(b) the nonzero mass at site $k$ hops to its nearest
neighbours. 

For the sites in the immediate neighbourhood of the origin (set $A$), the time
evolution equation for $P_k(m,t)$ is 
similar to that for set $B$ except that the contribution of the origin
needs to be taken into account separately. We have 
\bml
\bea
\frac{\partial P_{k}(m,t)}{\partial t}&=&
\sum_{\delta} {^\prime} \sum_{m^\prime=1}^m 
P_{k,k+\delta}(m-m^\prime,m^\prime,t) 
+ w \sum_{m^\prime \neq 0} P_{k,0}(m-1,m^\prime,t) 
-\sum_{\delta} {^\prime} \sum_{m^\prime \neq 0}
P_{k,k+\delta}(m,m^\prime,t) \no \\
&&- w \sum_{m^\prime \neq 0}P_{k,0}(m,m^\prime,t)
- \sum_{\delta} P_k(m,t)
\;\;,\;\;k \in A  \l{e21} \\ 
\frac{\partial P_{k}(0,t)}{\partial t}&=& \sum_{\delta} \sum_{m  \neq
0} P_{k}(m,t)-\sum_{\delta} {^\prime} \sum_{m {^\prime} \neq 0}
P_{k,k+\delta}(0,m^\prime,t) 
-w \sum_{m^\prime \neq 0} P_{k,0}(0,m^\prime,t)
\;\;,\;\;k \in A \;\;\;\;, \l{e22}  
\eea
\eml
where the primed sum denotes the sum over the nearest
neighbours, excluding the origin. 
The contribution of the origin is taken care of by the terms
with a coefficient $w$ which 
accounts for the gain or loss in mass via chipping from the origin. 

At the origin itself, the set of evolution equations obeyed are given by
\bml
\bea
\frac{\partial P_{0}(m,t)}{\partial t}&=&\sum_{\delta} [
\sum_{m^\prime=1}^{m} P_{0,0+\delta}(m-m^\prime,m^\prime,t)
+ w P_{0}(m+1,t)
- \sum_{m^\prime \neq 0} P_{0,0+\delta}(m,m^\prime,t)- w 
P_{0}(m,t) ] \l{e31} \\  
\frac{\partial P_{0}(0,t)}{\partial t}&=& \sum_{\delta} [ w P_{0}(1,t)
-\sum_{m^\prime \neq 0} P_{0,0+\delta}(0,m^\prime,t) ]
\;\;\;\;.\l{e32}   
\eea
\eml
In the gain term for nonzero mass, besides the convolution term, there
is an extra term due to the possibility of chipping off one extra
monomer with a rate $w$. The loss terms are similar to those discussed
for the sites in set $B$.

\subsection{The Phases}
\label{dap}

\begin{figure}
\begin{center}
\psfig{figure=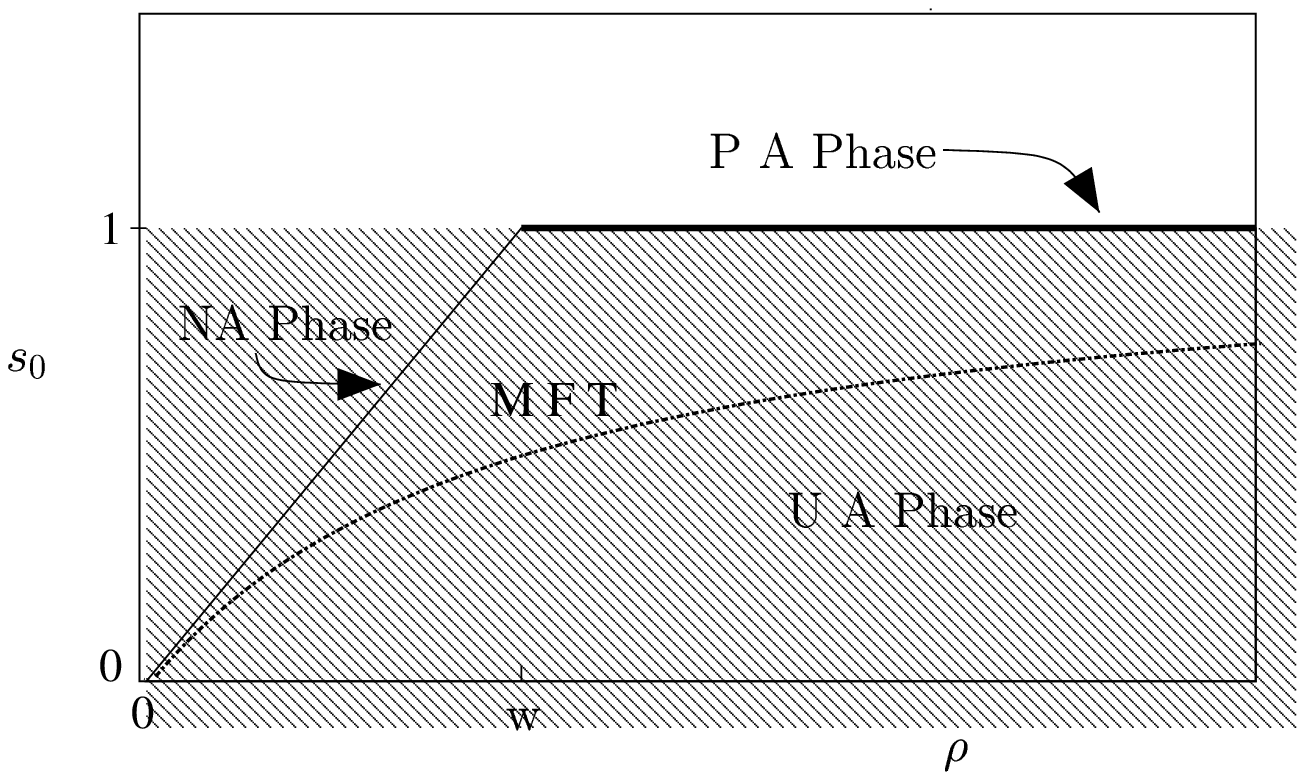,width=12cm,angle=0}
\caption{Plot of $s_0$ vs. $\rho$ to show the three possible
phases in the steady state of the aggregation model in the thermodynamic
limit. The system exists in (a) 
Pinned Aggregate phase when $s_0$ is one (bold horizontal line) (b)
Unpinned Aggregate phase when $s_0 < 1$ and not equal to $\rho/w$
(grey striped region) (c) Non Aggregate phase when $s_0$ is equal to
$\rho/w$ (dashed line). The smooth curve shows the MFT solution for
$s_0=\rho/(\rho+w)$ in the aggregation model.}
\label{s0}
\end{center}
\end{figure}

One can obtain useful information about the nature of the steady
state from 
the following simple analysis. In the steady state, 
the mass current into and out of any site must be equal, i.e.
\bml
\bea
\sum_{\delta} \;\av{m_k}&=&\sum_{\delta} \av{m_{k+\delta}} \;\;,\;\;k
\in B   \l{j1} \\ 
\sum_{\delta} \; \av{m_k}&=&w s_0 + \sum_{\delta} {^\prime} \av{m_{k+\delta}}
\;\;,\;\;k \in A  \l{j2}  \\
\sum_{\delta} w s_0 &=& \sum_{\delta} \av{m_{0+\delta}} \l{j3} \;\;\;\;,     
\eea
\eml
where $\av{m_k}$ is the average mass at site $k \neq 0$ and
$s_0=1-P_0(0)$ is the 
probability that the origin is occupied.  
The solution of the above equations gives $\langle m_k \rangle= w s_0$
for all $k \neq 0$. Thus the average mass on a site in the bulk is uniform
in spite of broken translational invariance.
Since the total mass of the system is conserved, the average
mass at the origin $\langle M_0 \rangle$ must satisfy
\be
\langle M_0 \rangle=\rho L^d - w s_0 (L^d-1) \l{cons} \;\;\;\;.
\ee
Using this constraint equation, one can deduce some characteristics of
the possible phases in the system as follows.
Since the LHS in Eq.(\ref{cons}) is non-negative and $s_0$ is bounded
above by one, the allowed domain for $s_0$ is constrained 
as shown in Fig.~\ref{s0}. Depending on the value of $s_0$ (and
hence $\av{M_0}$), the system can
exist in one of the three distinct phases described below.

\textbf{PA phase}: The system exists in this phase when $s_0$
is pinned to its maximum value $1$ for all $\rho > w$ (shown by the
bold line in Fig.~\ref{s0}). From the constraint equation Eq.(\ref{cons}), 
one deduces that the average mass at the origin grows
as $L^d$ giving rise to an infinite aggregate in the thermodynamic limit.
Thus the Pinned Aggregate phase is defined as that in which an
infinite aggregate is present at the origin with a probability one.
 
\textbf{UA phase}: The system exists in this phase for all values of $s_0$
lying in the shaded region in Fig.~\ref{s0}. This phase is
characterised by $s_0$ strictly 
less than $1$, and not equal to $\rho /w$. As in the PA phase, the average
mass at the origin grows as $L^d$ but $s_0$ is
not fixed and varies with $\rho$. Since $s_0 < 1$, the infinite
aggregate is present at the origin only for a finite fraction of the
time. For this reason, we refer to this phase as the Unpinned Aggregate phase.

\textbf{NA phase}: This phase is characterised by $s_0
\rightarrow \rho /w$ as $L \rightarrow \infty$ (shown by the solid line
in Fig.~\ref{s0}). Using Eq.(\ref{cons}), one immediately obtains 
$\av{M_0}/L^d  \rightarrow 0$ in the thermodynamic limit. Due to the
absence of an aggregate with mass of $O(L^d)$ at the origin, we call
this the Non Aggregate phase.

A more detailed study of the system shows that depending on the
dimension and drive, the system can either stay in
one of the three phases or else make transitions from 
one phase to another as $\rho$ is varied, keeping $w$ fixed. We have
solved for $\av{M_0}$ and $s_0$ in $d$ dimensions within a sitewise
inhomogeneous Mean Field Theory (Fig.~\ref{s0}) which predicts
that the system exists only in the UA phase 
in all dimensions, regardless of the bias. This 
prediction was tested numerically in several cases and seen to be
qualitatively correct except in the $1d$, biased case. 
\noindent In the latter case, 
there is a phase transition from the NA phase to the PA phase as one
crosses the critical line $\rho=w$ either by increasing $\rho$ with $w$
held fixed or by decreasing $w$ for fixed $\rho$.

\section{Unpinned Aggregate Phase: Mean Field Theory and Monte-Carlo
simulations}
\label{mft}

In this Section, we analyse the aggregation model within
the mean field approximation. We implement this approximation by
replacing the joint 
probability distribution $P_{k,k+\delta}(m,m^\prime,t)$ 
by the product $P_k(m,t)$ $P_{k+\delta}(m^\prime,t)$ of the single site
probability distributions.
We begin with the MFT analysis in arbitrary dimensions in
Section~\ref{mfth}. We will see that MFT predicts only the existence
of the UA phase in all dimensions, irrespective of bias. In
Section~\ref{mft1}, we study the UA phase in one dimension in more
detail. We discuss some limitations of this MFT and also
present some numerical results. We close this Section with a
qualitative discussion of the UA phase.

\subsection{MFT in arbitrary dimensions}
\label{mfth}

In this Section, we solve for $\av{M_0}$ and $s_0$ in arbitrary
dimensions and for all bias within MFT and show that it predicts the
occurrence of only the UA phase in all cases. 

In the steady state, the mass distribution at any site $k$ is
independent of time and is determined by setting the LHS of
Eqs.(\ref{e11})-(\ref{e32}) equal to zero. Further we replace
$P_{k,k+\delta}(m,m^\prime)$ by $P_k(m)$ $P_{k+\delta}(m^\prime)$ in
Eqs.(\ref{e11})-(\ref{e32}) to obtain the mean field equations for
$P_{k}(m)$. 
To study these equations, we define $Q_{k}(z)= \sum_{m=0}
P_{k}(m) z^m$, the Laplace transform of $P_{k}(m)$ with respect to
mass, and find that it 
obeys the following set of equations,
\bml
\bea
\sum_{\delta} Q_k &=& \sum_{\delta} \left[ 1+ Q_k \; (Q_{k+\delta}-1)
\right] \;\;,\;\;k \in B \l{mftI} \\
Q_k &=&\frac{\sum_{\delta} 1}{\sum_{\delta} 1+w s_0 (1-z)- \sum_{\delta}
{^\prime} (Q_{k+\delta} -1)} \;\;,\;\;k \in A \l{mft2} \\
Q_0&=&\frac{\sum_{\delta} w (1-z) (1-s_0)}{\sum_{\delta} [w (1-z) +z
(Q_{0+\delta} -1)]} \;\;\;\;. \l{mft3}
\eea
\eml
\begin{figure}
\begin{center}
\psfig{figure=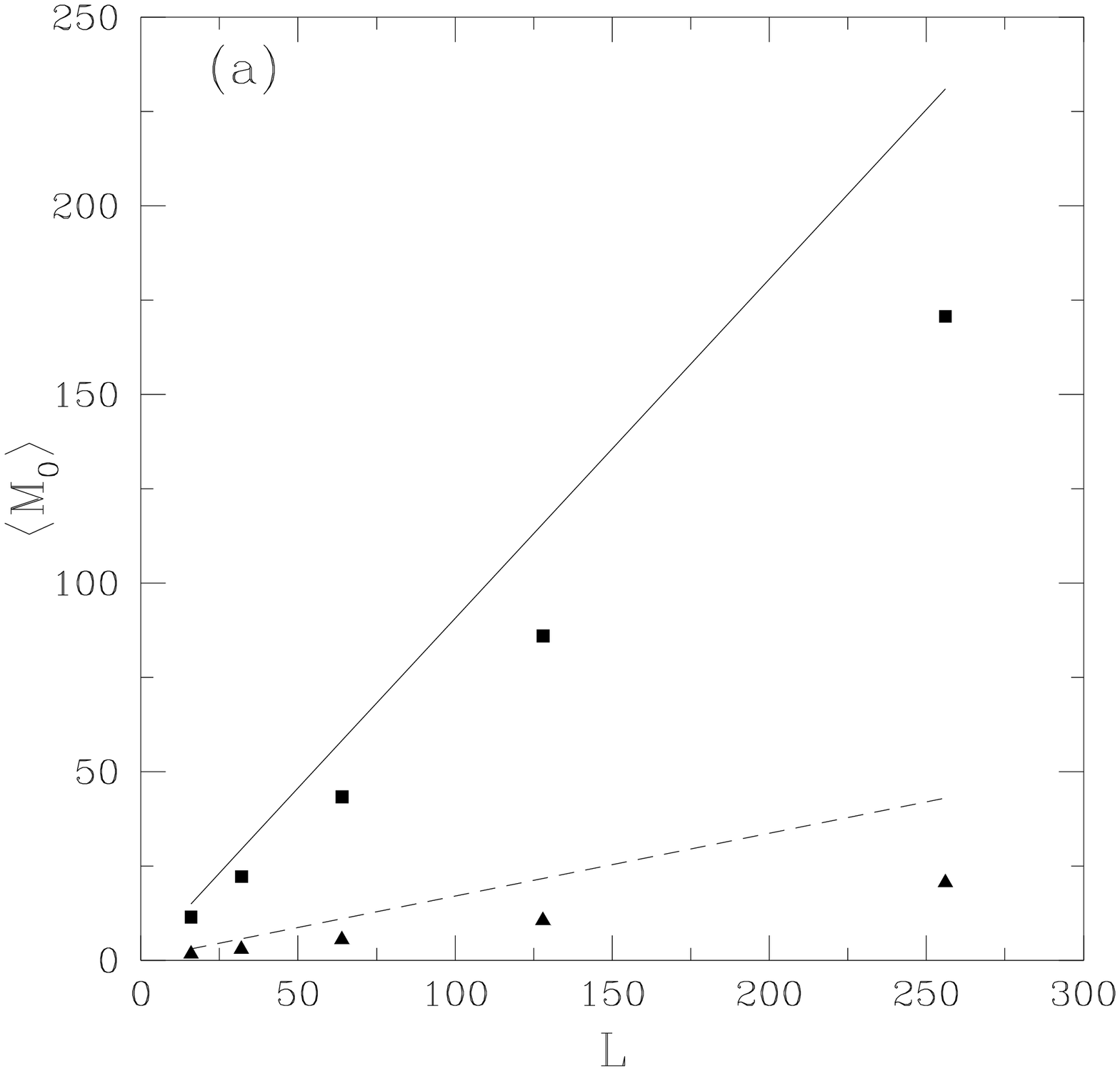,width=8cm,angle=0}
\psfig{figure=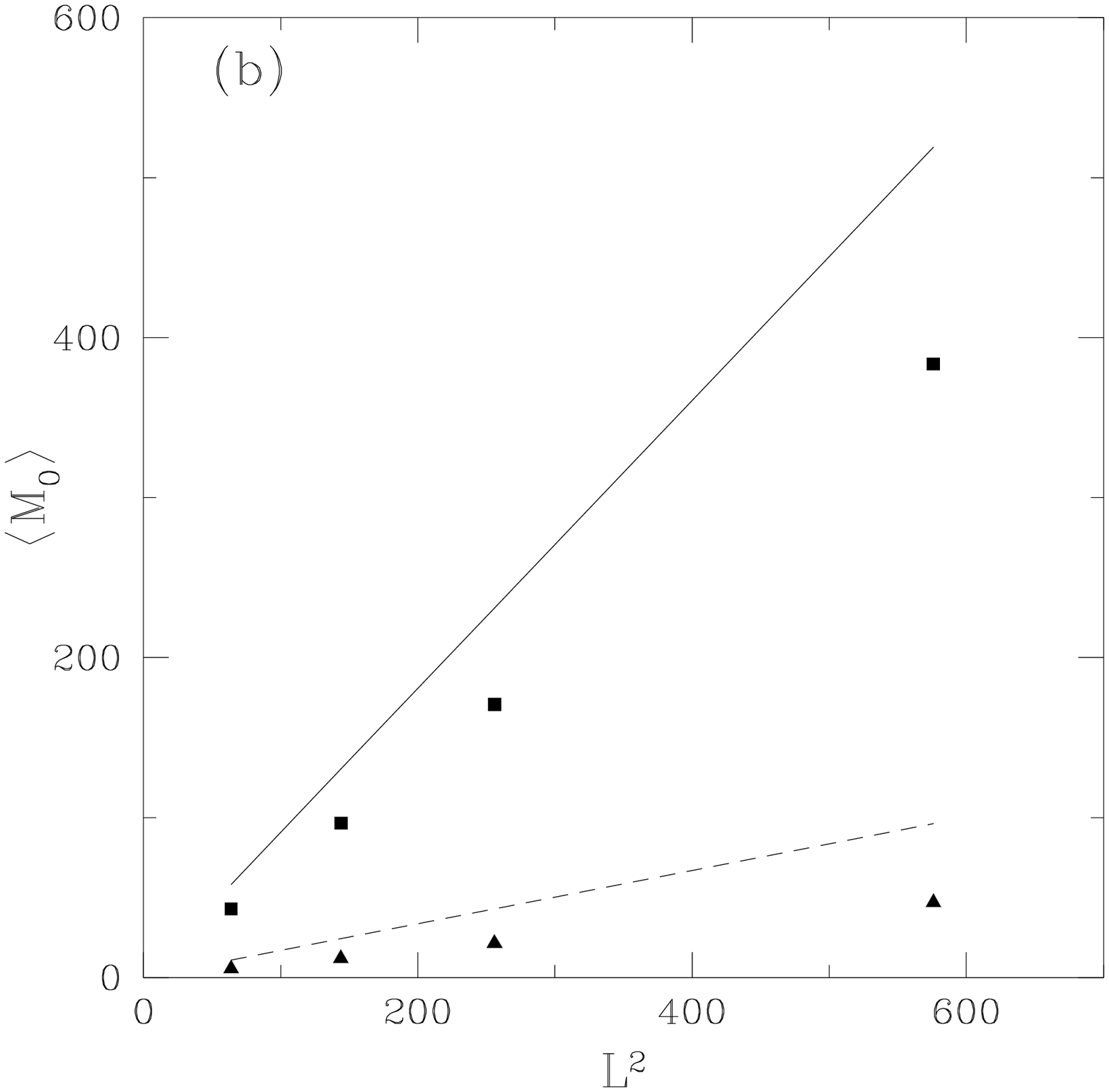,width=8cm,angle=0}
\caption{Plot of $\av{M_0}$ vs. $L^d$  
for the symmetric, aggregation model for (a) $d=1$ and (b) $d=2$. In each
case, the numerical result is plotted for two densities
$\rho=1.5$ (squares) and $\rho=0.5$ (triangles) alongwith the MFT 
prediction (solid line for $\rho=1.5$ and broken line for
$\rho=0.5$). Set of parameters used: w=1.} 
\label{1dsy}
\end{center}
\end{figure}

Since $P_{k}(m)$ is normalised to one, $Q_k(1)=1$ for all $k$ in the
above equations. The average mass in the bulk
$\langle m_k \rangle= {Q^\prime_k(z)}|_{z=1}$ still obeys
Eqs.(\ref{j1})-(\ref{j3}) which 
gives uniform mass in the bulk $\av{m}=w s_0$, and Eq.(\ref{cons}) still holds.

Although we could not solve the above set of nonlinear coupled
equations, we were able to compute 
two quantities of primary interest, namely, $\av{M_0}$ and $s_0$.
We begin by observing that the average mass at the origin can be
written in terms of the mean-squared mass at its nearest neighbours,
\be
\av{M_0}=\frac{\sum_{\delta} \left(2 w s_0 +
\av{m_{0+\delta}^2}-\av{m} \right)}{\sum_{\delta} 2 w (1-s_0)} \;\;\;\;,
\ee
where we have used that $\av{M_0}={Q^\prime_0(z)}|_{z=1}$ and
$\av{m_k^2}={Q^{\prime \prime}_k(z)}|_{z=1}+\av{m}$, $k \neq 0$. 

The mean-squared mass at sites other than the origin obeys the
following set of $\emph{linear}$ equations, 
\bml
\bea
\sum_{\delta} \av{m_k^2} &=& \sum_{\delta} \left( \av{m_{k+\delta}^2} +
2\av{m}^2 \right)  \;\;,\;\;k \in B \l{mk21} \\
\sum_{\delta} \av{m_k^2} &=& \sum_{\delta} {^\prime}
\left( \av{m_{k+\delta}^2} + 2\av{m}^2 \right) + 2 \av{m^2} + \av{m}
\;\;,\;\;k \in A \;\;\;\;. \l{mk22} 
\eea
\eml
It suffices to calculate $\sum_{\delta} \av{m_{0+\delta}^2}$ in order to
obtain $\av{M_0}$. Adding the above $L^d-1$ equations , one obtains 
\be
\sum_{\delta}\av{m_{0+\delta}^2}=\sum_{\delta} \left( 2 \av{m}^2
(L^d-1) + \av{m} \right) \;\;\;\;,
\ee
which further yields
\bea
\langle M_0 \rangle &=&\frac{ w s_0 +
 \av{m}^2 (L^d-1) }{w (1-s_0)} = \rho L^d -w s_0
(L^d-1) \;\;\;\;,
\eea
where the mass conservation equation Eq.(\ref{cons}) has been used in the
last identity. 
Solving for $s_0$ and $\langle M_0 \rangle$ in terms of $\rho, w$ and
$L$, we find
\bml
\bea
\langle M_0 \rangle &=& \frac{\rho L^d (1+\rho L^d)}{(1-w)+(\rho
+w)L^d} \l{M0d}\\ 
s_0 &=& \frac{\rho L^d}{(1-w)+(\rho+w) L^d}  \;\;\;\;. \l{s0d}
\eea
\eml
\begin{figure}
\begin{center}
\psfig{figure=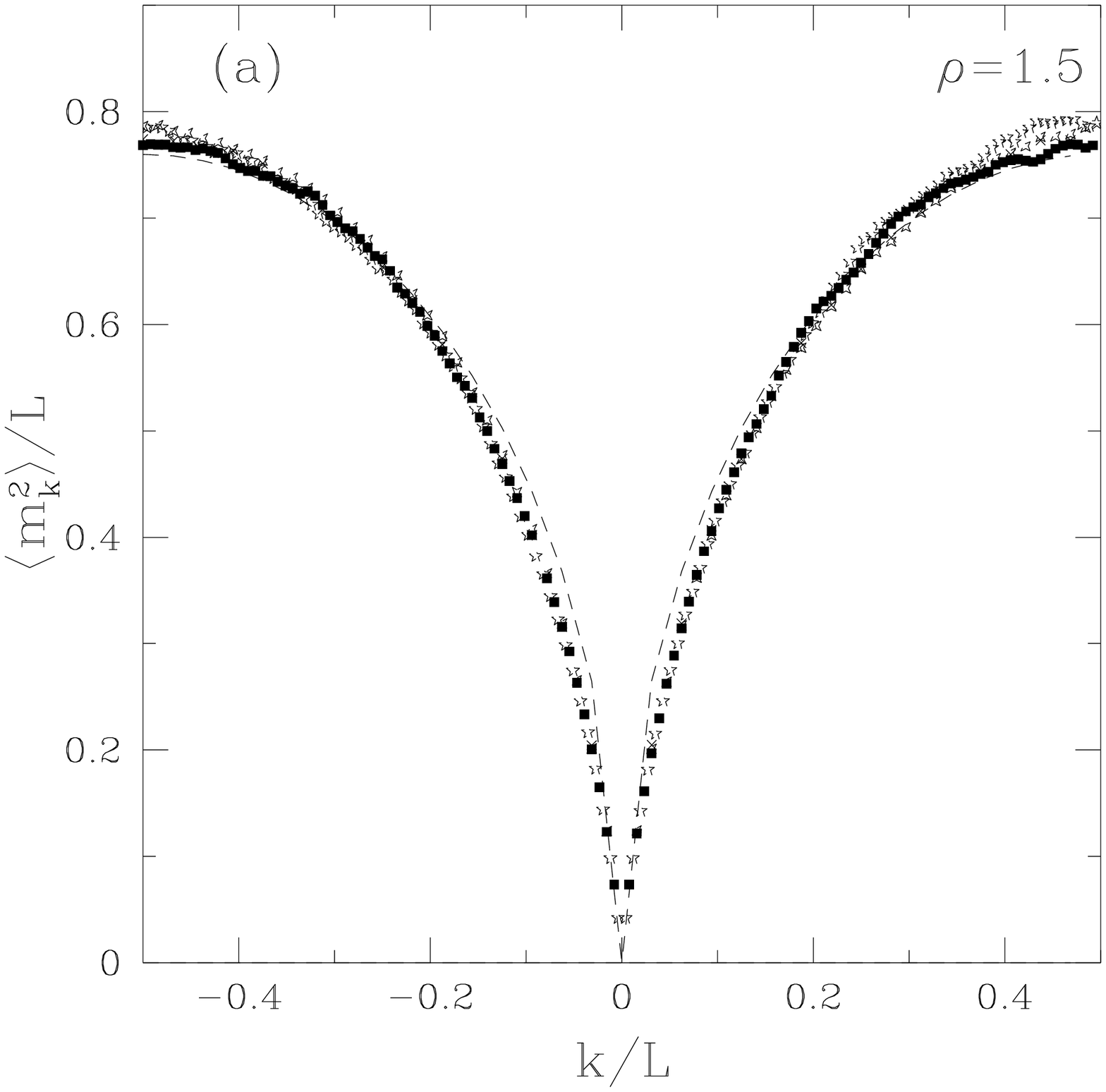,width=8cm,angle=0}
\psfig{figure=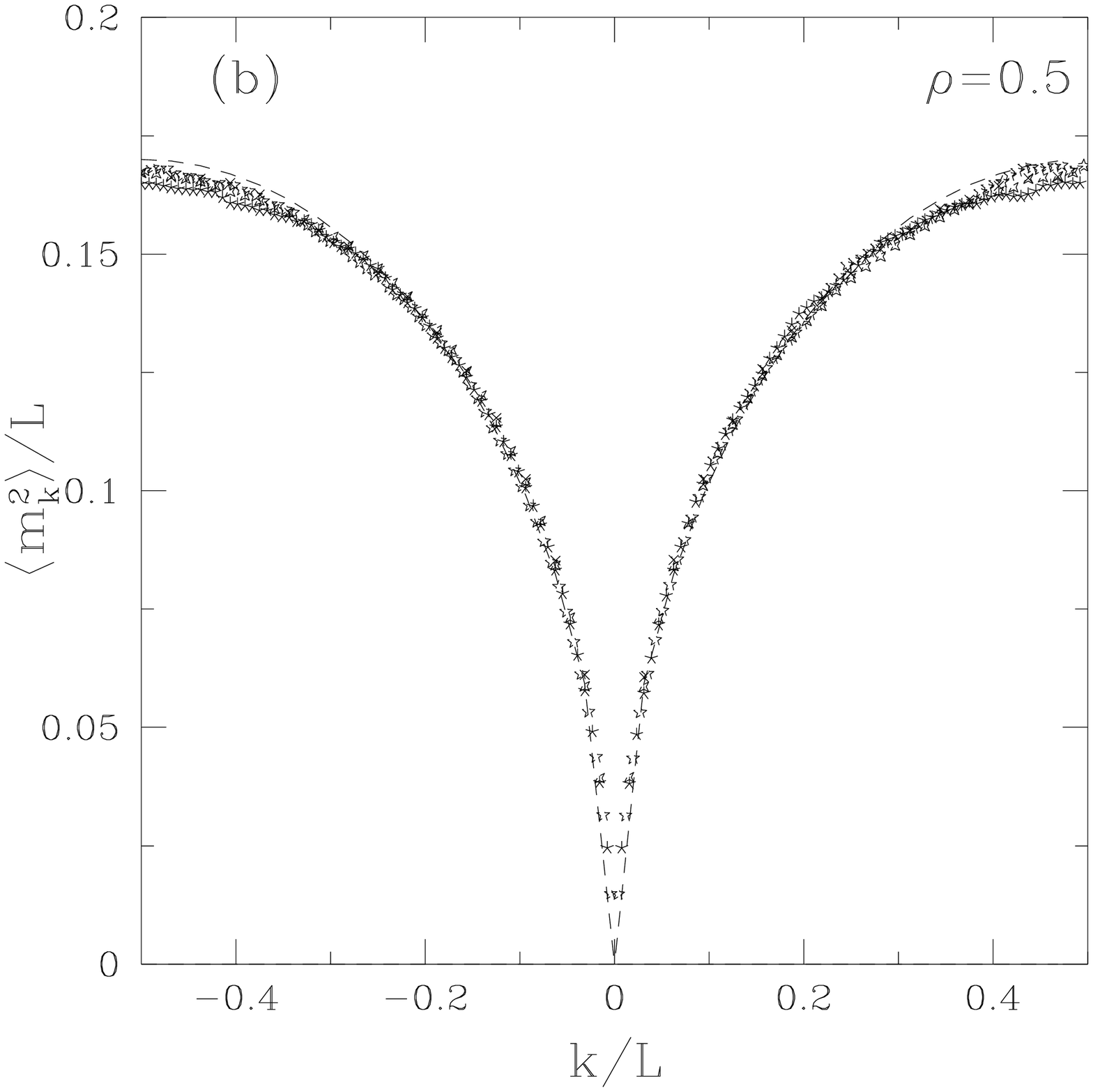,width=8cm,angle=0}
\caption{Data collapse for the mean squared mass at site $k \neq 0$ for
$1d$, symmetric aggregation model for $L=32, 64, 128, 256$ for (a) $\rho=1.5$
and (b) $\rho=0.5$. The numerical fit $y= A(\rho)\; \sqrt{x (1-x)}$
where $x=|k|/L$ is shown in broken line with $A(\rho=1.5)=1.52$ and
$A(\rho=0.5)=0.34$. Set of parameters used: w=1.}   
\label{mk21d}
\end{center}
\end{figure}
To lowest nonvanishing order in $L$, the above equations give
$\av{M_0} \sim L^d$ and $s_0 < 1$ which are the signatures of the UA
phase. Thus MFT predicts that the system exists only in the UA phase
for all $\rho$ and $w$ (see Fig.~\ref{s0}). Our numerical simulations
show that this prediction is true at least qualitatively in all cases
except in the $1d$, biased case. In Fig.~\ref{1dsy}, we show our
simulation results for $\av{M_0}$ in the absence of bias in $1d$ and
$2d$ plotted along with the MFT results,
and find qualitative agreement. We also measured $s_0$ for various
system sizes and densities, fixing $w=1$ and find that it is
independent of $L$ in 
accordance with Eq.(\ref{s0d}). For $\rho=0.5$, we found that
$s_0^{num}\simeq 0.42$ to be compared with 
$s_0^{MFT}\simeq 0.33$; for $\rho=1.5$, numerically $s_0^{num}\simeq 0.84$
whereas $s_0^{MFT}=0.6$.

\subsection{UA phase in one dimension}
\label{mft1}

In this Section, we analyse the UA phase in more detail in the $1d$,
unbiased case. We solve Eq.(\ref{mk21}) and Eq.(\ref{mk22}) for the 
mean-squared mass in the bulk in this case and find
that the MFT seems to violate the conservation law. We comment on this
limitation of MFT.
We also present numerical evidence which shows the simultaneous
presence of more than one infinite aggregate in the system.

For convenience, we will choose $L$ to be odd.  
The mean-squared mass at sites other than the origin obeys the
following equations,
\bml
\bea
2 \av{m_k^2} &=& \av{m_{k-1}^2} + \av{m_{k+1}^2} + 4 \av{m}^2
\;\;, \;\;|k| \geq 2  \\
2 \av{m_1^2} &=& \av{m_2^2} + \av{m} +4 \av{m}^2     \\
2 \av{m_{-1}^2} &=& \av{m_{-2}^2} + \av{m} +4 \av{m}^2 \;\;\;\;.
\eea
\eml
This set of equations can be solved and one obtains
\bea
\av{m_{k}^2}&=&2 \av{m}^2 (L-1)+ 2 \av{m}^2 (|k|-1) (L-|k|-1) +\av{m}
\;\;,\;\; k \neq 0 \;\;\;\;. \l{mk21ds} 
\eea
To leading order in $L$, the above result can be written in the
scaling form $\av{m_k^2}=L^2 f(|k|/L)$ where $f(x)=x(1-x)$ \cite{comment}.

The mean squared mass $\av{m_k^2}$ is actually a measure of the
typical mass at site $k$. To see this, note that we may define two
types of average over mass distributions at site $k$, namely, an
average $\langle ... \rangle$ over all mass occupations including
$m_k=0$, and an average $\langle \langle ... \rangle \rangle$ over
mass occupations excluding $m_k=0$. It is straightforward to see
that $\langle \langle ... \rangle \rangle = \langle ... \rangle/s_k$
where $s_k=1-P_k(0)$ is the probability that site $k$ is occupied. For
instance, 
the average mass $\av{m_k}$ 
\begin{figure}
\begin{center}
\psfig{figure=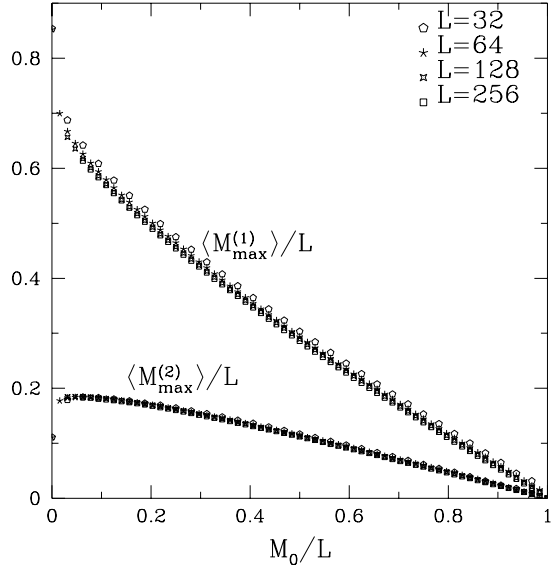,width=8cm,angle=0}
\caption{Data collapse for the conditional average masses $\av{M_{max}^{(1)}}$
and  $\av{M_{max}^{(2)}}$ of the clusters of largest 
and second largest mass in the bulk when the mass at the chipper is $M_0$ for
$1d$, symmetric aggregation model for $L=32, 64, 128, 256$. Set of
parameters used: w=1, $\rho=1$.}  
\label{2agg}
\end{center}
\end{figure}
is same on all sites while the typical
mass $\langle \langle m_k \rangle \rangle = \langle m_k \rangle/s_k$ will
depend on $k$.  Similarly, $\langle \langle m_k^2 \rangle \rangle$
is the square of the typical mass at site $k$, implying that 
$\av{m_k^2}=\langle\langle m_k^2 \rangle \rangle \;s_k$ may be
interpreted as a measure of the typical mass at site $k$.

We performed numerical simulations to test the mean field prediction for the
scaling form $\av{m_k^2}=L^2 f(|k|/L)$ where $f(x)=x(1-x)$.
As shown in Fig.~\ref{mk21d}, we obtain a data collapse 
with the scaling form $\av{m_k^2}=L g(|k|/L)$ with $g(x) \sim
\sqrt{x (1-x)}$ as the numerical fit for the scaling function. Since
$g(x) \sim \sqrt{x}$ for $x$ close to zero 
and is constant for $x \approx 1/2$, the typical mass scales as
$\sqrt{L}$ for $k$ close to the origin but as $L$ at sites diametrically
opposite to the origin. This points to the existence of an aggregate
in the bulk as well, consistent with the nomenclature Unpinned
Aggregate Phase. 

One notes that the MFT solution seems to violate the conservation law
since it predicts typical masses of $O(L^2)$ at sites situated $O(L)$
away from the chipper. On the other hand, one can also check that
Eq.(\ref{cons}) is true using the MFT results for $\av{M_0}$ and
$s_0$. Thus MFT seems to be able to describe the vicinity of the
chipper more correctly than the bulk. The reason for
this could be that the mass fluctuations about
the mean near the chipper are small due to fragmentation unlike 
those in the bulk where only the aggregation move operates. At a distance of
$O(L)$ away from the chipper, to a good approximation, one can neglect
the presence of the chipper. Then as all the mass resides only on one site,
the mass-mass 
correlation function at two different sites $\av{m_i m_j}$ is exactly
zero in the steady state, in strong contrast to the mean field approximation 
$\av{m_i m_j}=\av{m}^2$. Thus a more refined approximation is required
in the regions where the aggregation move dominates.

In Fig.~\ref{2agg}, we present the numerical evidence which indicates the
simultaneous presence of more than one infinite aggregate in the
system.
We measured the conditional average masses $\av{M_{max}^{(1)}}$ and
$\av{M_{max}^{(2)}}$ of the clusters of the largest and second largest mass 
in the bulk respectively, given there is a mass $M_0$ at the
chipper. We find that data collapse is obtained with the scaling forms
$\av{M_{max}^{(1)}}= L f_1(M_0/L)$ and $\av{M_{max}^{(2)}}= L 
f_2(M_0/L)$, where the 
scaling functions $f_1(x)$ and $f_2(x)$ decay almost linearly.  
Thus a
localised infinite aggregate at the chipper and one or more mobile
infinite aggregates in the bulk can be present at the same time; we would
expect more than one infinite aggregate to be present in higher
dimensions as well.

In any dimension $d$, far away from the chipper site, the state
resembles that in the absence of the chipper. In the latter case,
there is a single mobile infinite aggregate which is equally likely to
be present at any site with a probability $1/L^d$. Although this
mobile aggregate with mass of order $L^d$ arrives at the chipper 
infrequently, the 
probability of occupation $s_0$ of the chipper is of $O(1)$ -- this
enhancement occurs because mass can leave only one unit at a time, so
that it stays for a time of at least order $L^d$. It is implicit in
the above argument that an infinite aggregate can be formed in the
bulk before it hits the chipper. However, this fails to be true in the
$1d$, biased case which explains the absence of the UA phase in that
case.

\section{ASYMMETRIC SINGLE-CHIPPER AGGREGATION MODEL IN ONE DIMENSION}
\label{except}

The mean field prediction that the system exists in the UA phase fails
in $1d$, biased case. In the presence of a drift velocity, the
one-dimensional system undergoes a phase transition from the NA phase
to the PA phase as $\rho$ is increased, keeping $w$ fixed. 
This exceptional case is the subject of this Section. Although the
phase transition survives for all nonzero bias, we will 
only discuss the extreme case when the mass moves only forward
(i.e. infinite bias).   

The time required to form an aggregate with mass of $O(L)$ in the bulk is
$O(L^2)$. But in this case, due to ballistic motion, the mass clusters
return to 
the chipper in time of $O(L)$ ruling out the formation of an infinite
aggregate in the bulk. For small $w$, the sublinear mass arriving at
the chipper cannot 
leave it easily and is temporarily trapped giving rise to a localised infinite
aggregate (PA phase). As $w$ is increased, the mass leaves more
frequently rendering the trapping less
effective; also this chipped off mass cannot return before the chipper
gets empty so that an infinite aggregate cannot be sustained at the
chipper for large $w$ (NA phase). Thus there is a phase transition in
the $1d$, biased case as $w$ (or alternatively $\rho$) is varied. 

\begin{figure}
\begin{center}
\psfig{figure=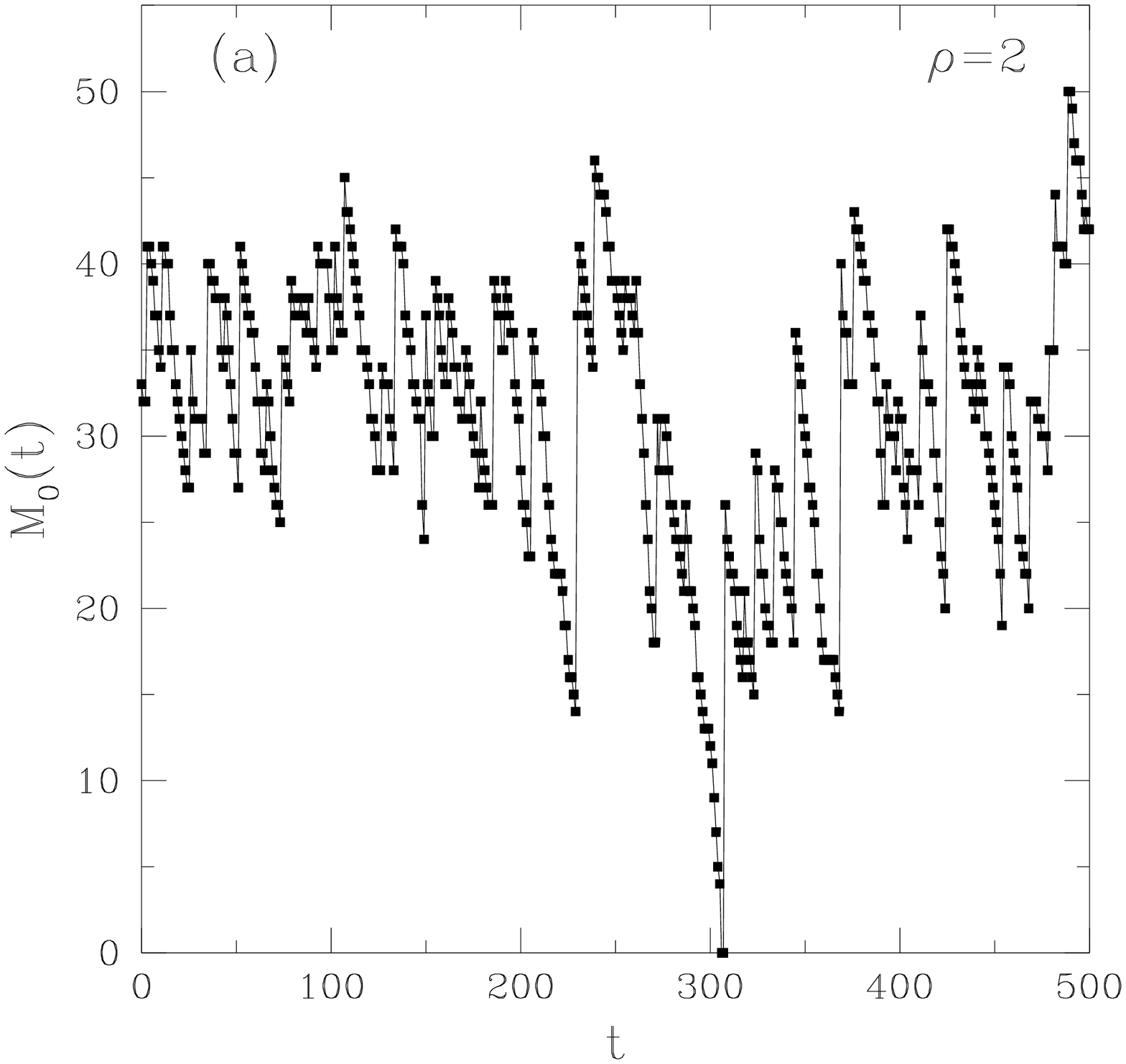,width=8cm,angle=0}
\psfig{figure=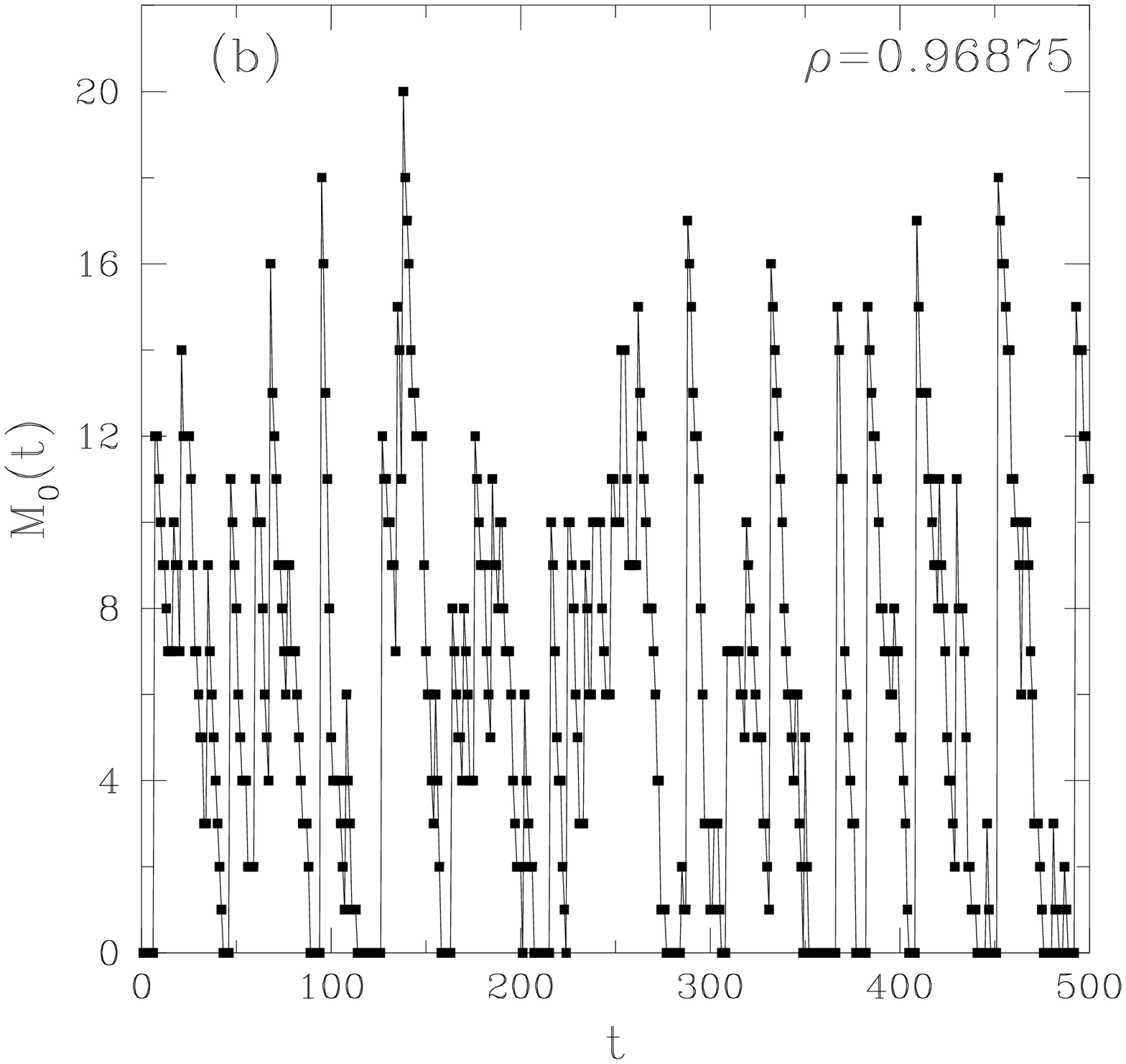,width=8cm,angle=0}
\psfig{figure=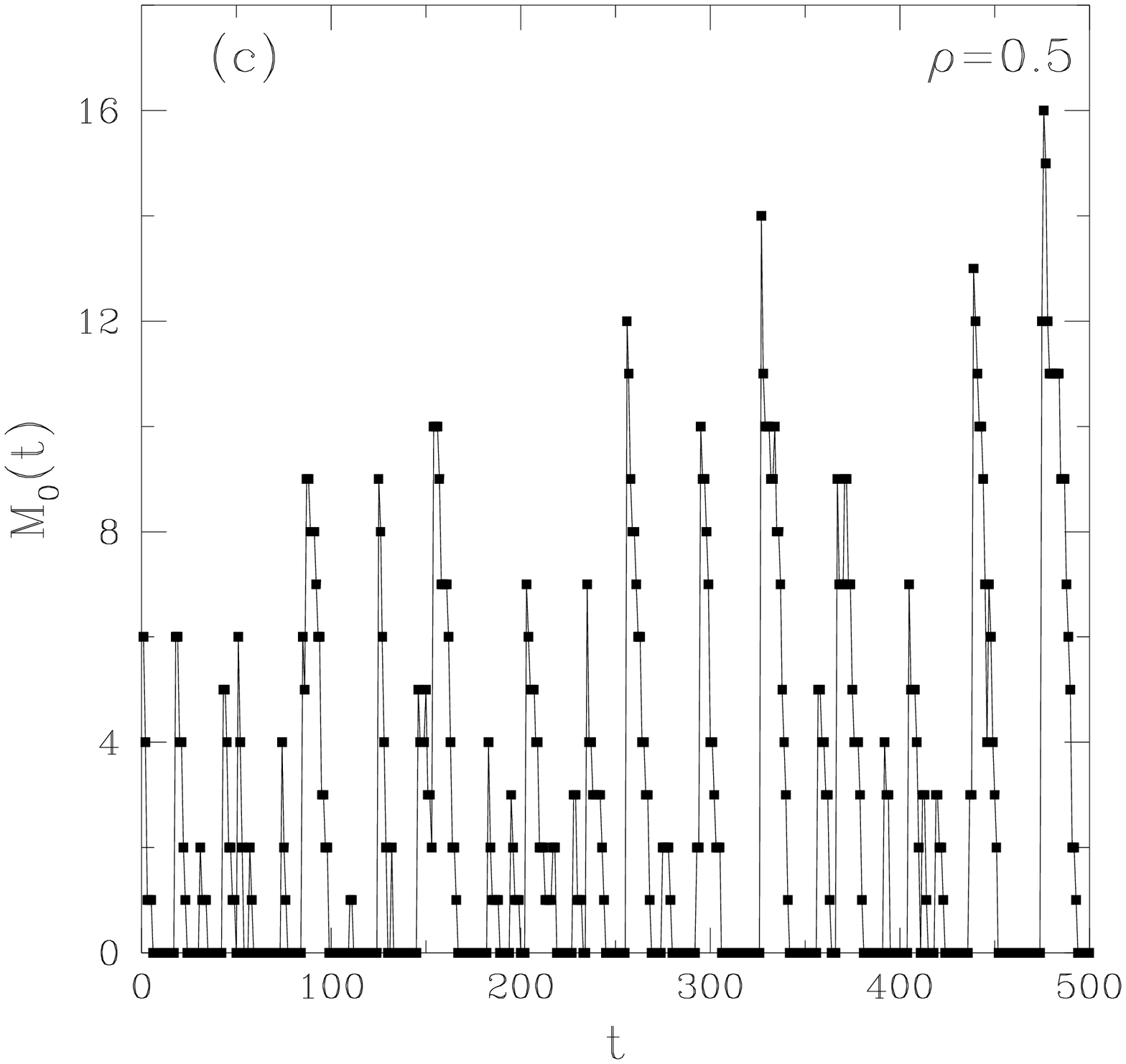,width=8cm,angle=0}
\caption{Instantaneous mass at the origin $M_{0}(t)$ vs. $t$ in the
steady state of $1d$, asymmetric aggregation model. (a) The PA
phase ($\rho=2$). (b) The critical point ($\rho=0.96875$). 
(c) The NA phase ($\rho=0.5$). Set of parameters used: L=32, w=1.}
\label{tser}
\end{center}
\end{figure}

The critical point is determined exactly to
be at $\rho_c=w$ in the thermodynamic limit by setting the  LHS to be zero
and $s_0=1$ in 
Eq.(\ref{cons}). For $\rho < \rho_c$, the system exists in the NA phase
which is characterised by $s_0=\rho/w$ and $\av{M_0}$ growing
sublinearly with $L$. For $\rho > \rho_c$, the system exists in the PA
phase in which $s_0$ is pinned to its maximum value $1$ and $\av{M_0}$
grows linearly with $L$ (see Fig.~\ref{s0}).
The probability $P_0(0)=1-s_0$ that the origin 
is empty serves as an order parameter. Our numerical simulations
indicate that to lowest order in $L$,  
\bea
1-s_0=P_0(0) \sim \left\{ \begin{array}{ll}
	    O(e^{-L/L_0})  & \;\;,\;\; \rho > w \\
	    O(L^{\beta-1}) & \;\;,\;\;\rho=w \\
	   \mbox{constant} & \;\;,\;\; \rho <w  \;\;\;\;.
	\end{array}
	\right.   \l{op}  
\eea
Since $P_0(0)$ varies continously from the
NA phase to PA phase as $\rho$ increases, this is a second order phase
transition. 

Before delving into details, we first present a
pictorial representation of the phases. Let us imagine monitoring the
instantaneous mass at 
the chipper site in the steady state. Figure~\ref{tser} shows the
time series for the mass at the origin as a function of time in the
two phases and at the critical point. The mass at the chipper
increases due to the mass input 
from the site behind the chipper, and decreases due to
fragmentation. Let $t_0$ and $t_1$ 
respectively denote the consecutive number of time steps during which the
chipper is empty and occupied. Then $s_0$ can
be related to $t_0$ and $t_1$ through $s_0^{-1}-1=\av{t_0}/\av{t_1}$. 
Depending on whether the ratio $\av{t_0}/\av{t_1}$ is zero or
not in the thermodynamic limit, $s_0$ is either pinned to $1$ (PA
phase) or strictly less than $1$ (NA phase). 
In the PA phase, $\av{t_1}/\av{t_0} \gg 1$ holds due to long cascades
of successive mass inputs, thus enabling the origin to maintain $s_0$ equal to
$1$. The long $t_1$ stretches (see Fig.~\ref{tser}(a)) enables the
chipper to build an  
aggregate with mass of order of system size. In contrast, in the
NA phase, $\av{t_0} \gtrsim \av{t_1}$ holds as 
depicted in Fig.~\ref{tser}(b) by the stretches of time during which
the origin is empty (due to absence of multiple inputs) 
which reduces $s_0$ from its maximum value $1$. Since $t_1$ is
typically not very long, the chipper cannot sustain an infinite
aggregate in NA phase.
Finally, at the critical point, there are multiple mass inputs to the
chipper but these cascades are not very long as shown in Fig.~\ref{tser}(c).
We now turn to a systematic discussion of each phase.

\subsection{The Pinned Aggregate Phase}
\label{pa}

Since the total mass in the system is conserved and the $L-1$ sites in
the bulk have nonzero average mass $\av{m}$ and sublinear fluctuations
about the mean mass (due to absence of a cluster of mass of order $L$
in the bulk as argued above), the probability $P_0(m)$ that the origin
has mass $m$ is $\sim 1/{\sqrt{L}}$ exp$(-{(m-m_0)}^2/L)$ where $m_0
\sim M-\av{m}L$ which
gives $P_0(0) \sim$ exp$(-L)$.
Thus in this phase, the chipper site is occupied with
probability one in the thermodynamic limit so that it
acts as a reservoir of particles for the bulk. One can think of
the system as a semi-infinite, one dimensional lattice with
a perfect, localised source at the origin injecting monomers into it
at the rate $w$.

The problem of aggregation in the presence of such a   
source has been considered previously as well. Some properties were
studied in \cite{derrida} using the technique of interparticle
distribution function (IPDF) introduced in
\cite{ipdf}, and in \cite{redner} by mapping this $1d$ problem to a
bounded random walk in $2d$. Here we 
calculate the steady state mass distribution at site $k$ denoted by
$P_k(m)$ by the generating function method.
We define the $r$-point characteristic function for site $k$ at time $t$, 
$Z_r^{(k)}(\lambda, t)=\langle e^{-\lambda \sum_{j=k}^{k+r-1} m_j(t)}
\rangle$. 
The time evolution equations obeyed by $Z_r^{(k)}(\lambda, t)$ with a
perfect, localised source at $k=0$ are given by 
\bml
\bea
\frac{\partial {Z_r^{(k)}}}{\partial  t}&=&
Z_{r+1}^{(k-1)}+Z_{r-1}^{(k)}-2 Z_r^{(k)}
\;\;,\;k \neq 1 \;\;,\;\; r \neq 0 \l{bulk} \\ 
\frac{\partial {Z_r^{(1)}}}{\partial  t}&=&
Z_{r-1}^{(1)}+ (w e^{-\lambda}-
1-w) Z_r^{(1)}\;\;,\;\; r \neq 0 \l{first}
\eea
\eml
with the boundary condition $Z_0^{(k)}(\lambda,t)=1$ for all $k > 0$.
In the steady state, $Z_r^{(k)}(\lambda, t)$ is independent of $t$. 

We need to solve for $Z_1^{(k)}(\lambda)$ which is the Laplace transform
of $P_k(m)$ with respect to $m$.
One can easily solve for $Z_1^{(1)}(\lambda)$ using 
Eq.(\ref{first}) and the boundary condition $Z_0^{(1)}=1$ which 
on taking the inverse Laplace transform for large $m$ gives
$P_1(m)=e^{-m/w}/w$. Thus the mass distribution at the
first site decays exponentially. 
Now to find the probability distribution in the bulk, we define 
\be
H(x,y,\lambda)=\sum_{k=2,...}\sum_{r=1,...} Z_r^{(k)}(\lambda) x^k y^r
\;\;\;\;. 
\ee
\begin{figure}
\begin{center}
\psfig{figure=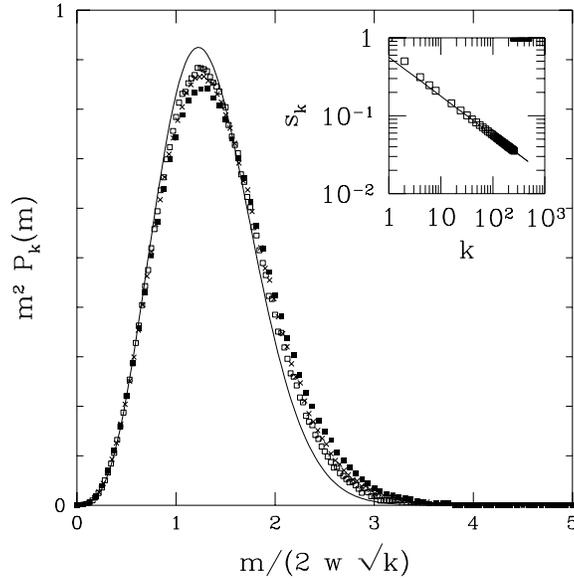,width=8cm,angle=0}
\caption{Data collapse for the unnormalised probability distribution of mass at
site $k=64$ (solid squares), $128$ (crosses), $256$ (squares) in the PA
phase plotted against the analytical 
result. Inset: log-log plot of probability distribution for nonzero 
mass at site $k$ plotted against the analytical result. Set of
parameters used: $L=256$, $\rho=2$, $w=1$.} 
\label{aggprob}
\end{center}
\end{figure}
Using equations Eq.(\ref{bulk}) and Eq.(\ref{first}), one obtains
\bea
H(x,y,\lambda)&=&\frac{xy}{y^2-2 y+x} \left[G(x,\lambda)-x y
\left(\frac{1}{1-x}  \right. \right. \left. \left.  +\frac{f^2}{1-fy} \right) \right] \;\;\;\;,
\eea
where $1/f=1+w-we^{-\lambda}$ and $G(x,\lambda)=\sum_{k=2,...}
Z_1^{(k)}(\lambda) x^k$. Clearly, the inverse Laplace transform of
$G(x,\lambda)$ 
w.r.t. $\lambda$ and $x$ gives $P_k(m)$ for $k >1$. 
To calculate $G(x,\lambda)$, we use the same method as in
\cite{beadpack}. We first note that the denominator of
$H(x,y,\lambda)$ has roots at  
$y_{\pm}=1\pm \sqrt{1-x}$. The root at $y=y_{-}$ lies inside the unit
circle which in the real space (i.e. $r$-space) gives an
exponentially increasing solution which is disallowed since the inverse
Laplace transform of $H(x,y,\lambda)$ gives a probability, which is always
bounded above. To avoid this pathological solution, we demand
that $y=y_{-}$ is a zero of the numerator as well, i.e.
\be
G(x,\lambda)=x y_{-} \left( \frac{1}{1-x}+\frac{f^2}{1-f y_{-}} 
\right)  \;\;\;\;.
\ee
For $\lambda \rightarrow 0$ and $x \rightarrow 1$, the inverse Laplace
transform can be easily found. We find that $P_k(m)$ has a scaling
form $\phi(u)/m^2$ where $u=m/2 w \sqrt{k}$ and
the scaling function is  
\be
\phi(u)= \frac{4 w}{\sqrt{\pi}} u^3 e^{-u^2} \;\;\;\;.
\ee

One can also obtain the probability distribution $P_k(0)$ of zero mass at site
$k$ by solving for $G(x,\lambda)$
when $\lambda \rightarrow \infty$ and $x \rightarrow 1$. To leading order
in $k$, one obtains $P_k(0)=1-s_{k}=1-1/\sqrt{\pi k}$.
We tested our calculations against the numerical simulations performed on
the single-chipper model and found reasonably good agreement (see  
Fig.~\ref{aggprob}). Thus the typical mass at site $k$ grows as $\sqrt{k}$. 

We now calculate the typical spacing between the masses using the method
of IPDF \cite{ipdf}.
We define $E_k(r,t)$ as the probability that the sites $k$ to $k+r$ are
empty (including $k$ and $k+r$). Then $E_k(r,t)$ satisfies the
following equations,
\bml 
\bea
\frac{\partial E_k(r,t)}{\partial t} &=& E_k(r-1,t)-2 E_k(r,t)+ E_{k-1}(r+1,t)
\;\;,\;\;k \neq 1  \\
\frac{\partial E_1(r,t)}{\partial t} &=& E_1(r-1,t)-(1+w) E_1(r,t)  \;\;\;\;,
\eea
\eml
with the boundary condition $E_{k}(-1,t)=1$ for all $k >0$.
In the steady state, $E_k(r,t)$ is independent of $t$ and one can
write the above equation in terms of
$F_{1}(z,y)=\sum_{k=2,...}\sum_{r=0,...} E_k(r) z^k \; y^r$ and
$R(z,r)=\sum_{k=2,...} E_k(r) z^k$ , $r=0,1,...$. 
We find 
\bea
F_{1}(z,y)&=&\frac{y}{y^2-2y+z} \left[\frac{z
R(z,0)}{y}-\frac{z^2}{(1+w)(1+w-y)} \right.\left. -\frac{z^2}{1-z} \right]  \;\;\;\;.
\eea
With the same reasoning as in the previous calculation, we demand that
the numerator of 
$F_{1}(z,y)$ evaluated at $y_{-}=1-\sqrt{1-z}$ be zero. This condition gives
\be
R(z,0)=\left[\frac{1}{1-z}+\frac{1}{(1+w)(1+w-y_{-})} \right] z \;
y_{-} \;\;\;\;,
\ee
which on inverting the Laplace transform gives the probability
$E_k(0)$ which is same as $P_k(0)$. One can check that the result for
$P_k(0)$ quoted above is reproduced.
Substituting $R(z,0)$ in $F_{1}(z,y)$, we obtain 
\be
F_1(z,y)=\frac{-z^2}{y-y_{+}} \left[ \frac{1}{1-z} +
\frac{1}{(1+w-y)(1+w-y_{-})} \right]   \;\;\;\;,
\ee
where $y_{+}=1+\sqrt{1+z}$.

We further define $D_k(r)$ as the probability that both sites 
$k$ and $k+r$ are occupied but no sites in between. Then
\bea
D_k(r)&=& E_{k+1}(r-2)- E_{k+1}(r-1)-E_{k}(r-1) +E_{k}(r) \;\;,\;\;k \neq 1, r \neq 0 \;\;\;\;.
\eea
Defining the Laplace transform of $D_k(r)$ with respect to $k$ and $r$
as $F_{2}(z,y)=\sum_{k=2,...}\sum_{r=0,...} D_k(r) z^k y^r$, we obtain
\bea
F_2(z,y) &=& \frac{1}{z} \left( R(z,-2)+ \frac{y z^2}{1-z} \right) - \frac{z
(1+z)}{1-z}+ F_{1}(z,y) 
\left( \frac{y^2}{z} \right. + \left. (1-y) (1+\frac{1}{z})  \right)
\\ \no
&&+ z (1-y) \left( 1+\frac{y}{2-y} 
\right. \left. \left\{ 1+ \frac{1}{(1+w)(1+w-y)}
\right\} \right) \;\;\;\;,
\eea
where $F_{1}(z,y)$ was calculated above.

Then the typical empty space $\av{r_k}$ in front of site $k$ can be obtained by
taking the inverse Laplace transform of
$(\partial F_{2}/\partial y)|_{y=1}$ with respect to $z$. One finds 
\be
\av{r_k}=1+ \frac{2}{\sqrt{\pi k}}- w e^{w^2 k} \mbox{erfc}(w
\sqrt{k}) \;\;\;\;,
\ee
which for large $k$ implies that the typical empty space
$\av{\av{r_k}}=\av{r_k}/s_k$ in front of occupied site $k$ varies as
$\sqrt{k}$. 

Thus, in the PA phase, the mass in the bulk is distributed in
$\sqrt{L}$ clusters with typical mass $\sqrt{k}$ at
site $k$, and an empty stretch of length $\sqrt{k}$ in front of it.

\subsection{The Non Aggregate Phase}
\label{na}

As discussed earlier, in the NA phase, $s_0$ is strictly less
than $1$ due to substantial time stretches of typical length
$\av{t_0}$ during which the origin is not occupied
(Fig.~\ref{tser}). Thus the  
origin does not act as a perfect source unlike in the PA phase and
we could not calculate the mass distribution in this phase. However,
one can obtain useful information about the 
nature of this phase by simple scaling arguments.
We begin by arguing that $\av{M_0}$ grows sublinearly with $L$ and
then provide numerical evidence for the absence of an infinite
aggregate (i.e. mass proportional to $L$) in the
bulk, thus concluding that there is no infinite aggregate anywhere.

\begin{figure}
\begin{center}
\psfig{figure=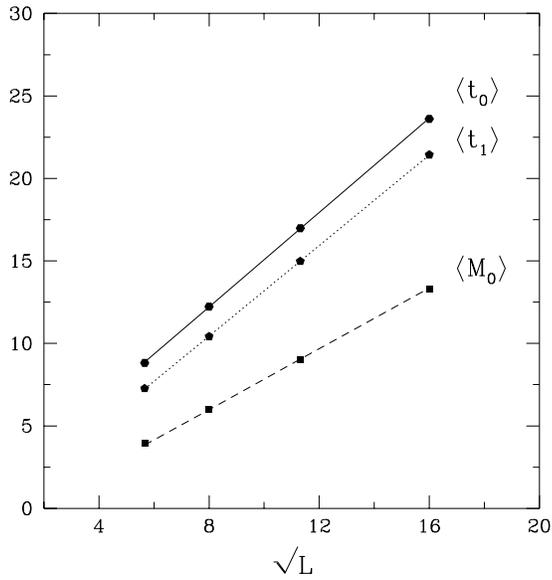,width=8cm,angle=0}
\caption{Plot of $\av{M_0}$, $\av{t_0}$ and $\av{t_1}$ as a function
of $\sqrt{L}$ in the NA phase. Since all the three curves are linear,
it follows that $\av{M_0}=a L^{\beta}+b$, $\av{t_0}=c_0
L^{\gamma}+d_0$ and $\av{t_1}=c_1 L^{\eta}+d_1$ with $\beta=\gamma=\eta=0.5$. 
Set of parameters used: $\rho=1$, $w=2$.} 
\label{naexpo}
\end{center}
\end{figure}

Monte Carlo simulations indicate that $\av{M_0}$, $\av{t_0}$ and
$\av{t_1}$ vary as a 
power law in $L$ with a constant as a next order
correction (see Fig.~\ref{naexpo}). Therefore, we assume 
$\av{M_0}=a L^{\beta}+b$, $\av{t_0}=c_0 L^\gamma+d_0$ and
$\av{t_1}=c_1 L^\eta+d_1$.   
Solving for $s_0$ using Eq.(\ref{cons}), we obtain
\be
s_0=\frac{\rho}{w} +O \left( \frac{1}{L^{1-\beta}} \right) \l{b1} \;\;\;\;.
\ee
Then since the lowest order term for $s_0$ is a constant and
$s_0^{-1}-1=\av{t_0}/\av{t_1}$, one obtains  
$\gamma=\eta$. Using this identity and retaining terms to lowest order in $L$ 
we further obtain
\be
s_0=\frac{c_1}{c_0+c_1}+ O \left( \frac{1}{L^\gamma} \right) \l{b2} \;\;\;\;.
\ee
Comparison of Eq.(\ref{b1}) and Eq.(\ref{b2}) yields $\beta+\gamma=1$.
\begin{figure}
\begin{center}
\psfig{figure=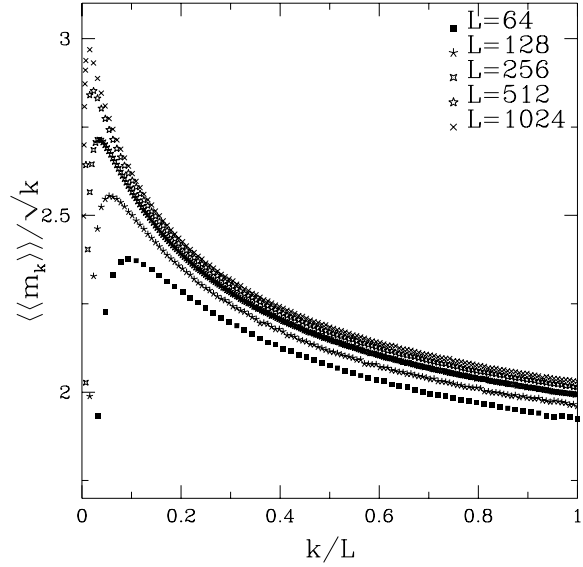,width=8cm,angle=0}
\caption{Data collapse for the typical mass at site $k \neq 0$ in the NA
phase for $L=64, 128, 256, 512, 1024$. Set of parameters used:
$\rho=1$, $w=2$. In this figure, all the sites are labelled by
positive integers in the clockwise direction with the chipper at the
origin.} 
\label{naco}
\end{center}
\end{figure}

Now we consider the typical mass $\av{\av{m_k}}$ in the bulk where the
site index $k$ is labelled by positive integers in the clockwise
direction with the chipper at the origin. Numerically,
$\av{\av{m_k}}$ is observed to obey the scaling form, 
$\av{\av{m_k}}= \sqrt{k} f \left( k/L \right)$ as shown in the data
collapse in Fig.~\ref{naco}. The
scaling function is a slowly varying function, which gives sublinear
mass everywhere in the bulk. Thus in the NA phase, the mass is
distributed in clusters of typical mass $\sim O(\sqrt{L})$ for $k \sim
O(L)$. 

Since the mass at the site just behind the chipper is of $O(\sqrt{L})$ and
as one can see in Fig.~\ref{tser}, the number of mass inputs to
the chipper is typically of $O(1)$, it follows that $\av{t_1} \sim
O(\sqrt{L})$ so that  
$\eta=0.5$, which further gives $\beta=\gamma=\eta=0.5$ due to the two
scaling relations above. These exponent values are checked numerically,
as shown in Fig.~\ref{naexpo}. 

\subsection{The Critical Point}
\label{cp}

We studied the critical point mainly numerically by studying the $L$
dependence of $\av{M_0}$, $\av{t_0}$ and $\av{t_1}$. 
Assuming a power law dependence for $\av{M_0}  \sim  L^{\beta}$,
$\av{t_0} \sim  L^{\gamma}$ and $\av{t_1}  \sim  L^{\eta}$
, a naive best fit in the range $L=32$ to $2048$ 
gives $\beta \simeq 0.62$, $\gamma \simeq 0.38$ 
and $\eta \simeq 0.80$. However, the effective exponent calculated using the
successive ratios of $L$ used in the simulations shows systematic trends: 
$\beta$ decreases as $L$ increases while both $\gamma$ and $\eta$
increase with $L$.
\begin{figure}
\begin{center}
\psfig{figure=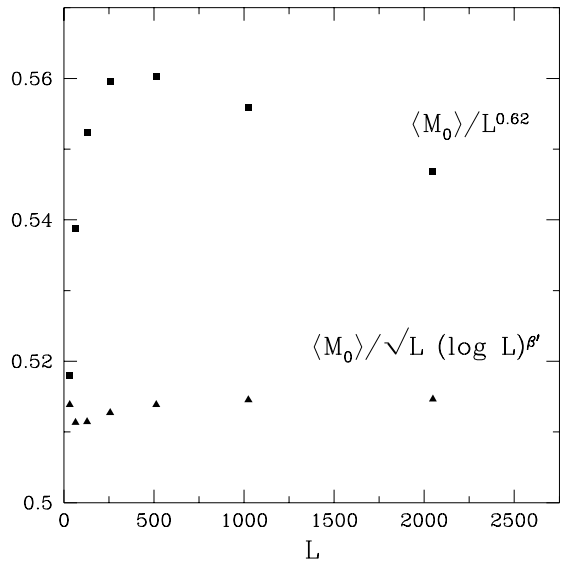,width=8cm,angle=0}
\caption{Plot of $\av{M_0}/ L^{0.62}$ (squares) and 
$\av{M_0}/\sqrt{L} \; (\log L)^{\beta^{\prime}},\beta^{\prime} \simeq 0.7$
(triangles) as a function of $L$ at the critical point in the $1d$,
biased case. The variation of $\av{M_0}$ as $\sqrt{L}$ with multiplicative
logarithmic correction seems to be a better fit than the fit to a pure
power law. The plot of $\av{M_0}/L^{0.62}$ vs. $L$ has been scaled by
a constant factor. Set of parameters used: $\rho=1, w=\rho L/(L-1)$.}
\label{crit1}
\end{center}
\end{figure}  

\begin{figure}
\begin{center}
\psfig{figure=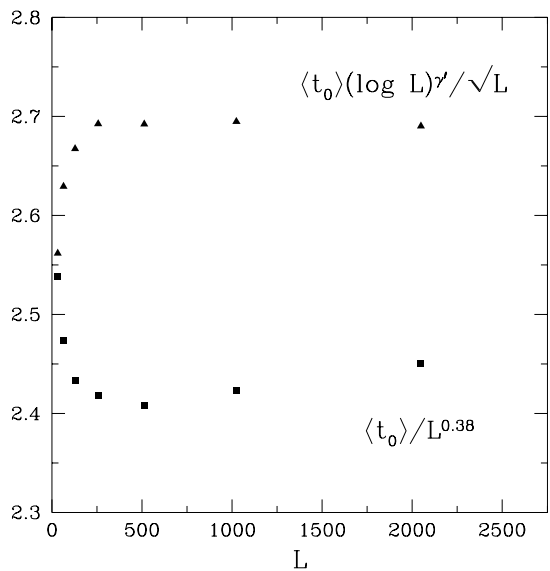,width=8cm,angle=0}
\caption{Plot of $\av{t_0}/ L^{0.38}$ (squares) and 
$\av{t_0} (\log L)^{\gamma^{\prime}}/\sqrt{L},\gamma^{\prime} \simeq 0.74$
(triangles) as a function of $L$ at the critical point in the $1d$,
biased case. The variation of $\av{t_0}$ as $\sqrt{L}$ with multiplicative
logarithmic correction seems to be a better fit than the fit to a pure
power law. The plot of $\av{t_0}/L^{0.38}$ vs. $L$ has been scaled by
a constant factor. Set of parameters used: $\rho=1, w=\rho L/(L-1)$.}
\label{crit2}
\end{center}
\end{figure}
\noindent A good fit is obtained if one allows for logarithmic
corrections in the power 
laws for $\av{M_0}$ and $\av{t_0}$ with $\beta=\gamma=1/2$ (see
Fig.~\ref{crit1} and Fig.~\ref{crit2}). This indicates that both
$\av{M_0}$ and $\av{t_0}$ vary as $\sqrt{L}$ with multiplicative
logarithmic corrections. 
Using Eq.(\ref{cons}) and the
condition of criticality, one can easily show that 
$\eta+\beta -\gamma =1$. Using $\beta=\gamma=1/2$ as suggested by the
discussion above, we obtain $\eta=1$. The above quoted values for the
exponents $\beta$ and $\gamma$ are consistent with the scaling relation
$\beta + \gamma=1$ which
seems to be true in the NA and the PA phase as well.
In the NA phase, we have seen in Section~\ref{na} that this scaling
relation holds. In the PA phase, we found numerically that $\av{t_0}$ is
independent of $L$ in the thermodynamic limit implying that
$\gamma=0$; further since $\beta=1$, it follows that the scaling relation
$\beta+\gamma=1$ holds in this phase also. 

We have also numerically studied $P(t_1)$ which is the probability
that the origin is occupied for $t_1$ consecutive time steps. At the critical
point, this 
distribution shows two peaks -- a broad peak occurs at the typical time scale
$t_1 \sim L^{\beta}$ while there is a narrow peak
at $t_1= N_{p}/w$. We have not been able to reliably separate
out these two contributions to $P(t_1)$ but since the scaling relation
yields $\eta=1$, one may expect that
the second peak dominates the asymptotic value of $\eta$.

\section{Single-Chipper Free particle model} 
\label{free}

\subsection{The Model}
\label{scfm}

In this Section, we study the steady state of a model of
noninteracting particles in which 
a localised infinite aggregate is formed solely due to disorder. 
In this model the particles diffuse freely in the bulk except at
certain, quenched sites referred to as the chipper sites. It can be shown
that this system exhibits a phase 
transition from the NA phase to the PA phase (as defined in
Section~\ref{dap}) as the 
particle density is increased in all dimensions for all bias. In this
Section, we will
demonstrate this result in the presence of a single chipper. We will
see that even a single defect is capable of inducing a phase
transition in all dimensions, unlike in the aggregation model. We will
discuss the solution with an extensive number of chipper sites in
Section~\ref{extens}.

Our model is defined on a $d$-dimensional hypercubic lattice with
periodic boundary conditions on which we consider the biased and unbiased
diffusion of a conserved number of particles.
At any site except the chipper, each particle attempts
to hop out at rate $1$. Since the particles do not interact, the
hopping rate out of site $k$ occupied by $m$ particles is 
$m$. At the chipper site, assumed to be at the origin, the hopping
rate is a constant $w$, independent of the number of particles. Thus,
the rate $u_k(m)$ at which a particle leaves site $k$ which has $m$
particles is given by
\be
u_k(m)= w \;\delta_{k,0} + m (1-\delta_{k,0}) \;\;,\;\;m \neq 0
\l{rates} \;\;\;\;.
\ee

In the fully asymmetric, $1d$ case, the model described above can be
mapped onto a traffic 
model on a one lane road with no overtaking allowed. We represent each
site in this model as a car and each particle as a  
vacant site. Then a system of $M$ free particles on a lattice of size
$L$ maps onto a system of $L$ particles with hard core interactions on
a lattice of size $L+M$. 
In this new representation, the special car (corresponding to the
chipper site) moves with a constant rate, irrespective of the headway
in front of it and rest of the cars (sites other than the chipper) move
with a rate proportional to the headway in front of it. 
Note that the $\emph{sitewise}$ disorder in the free particle model
corresponds to $\emph{particlewise}$ disorder in the 
traffic model. 

\subsection{Phase Transition in arbitrary dimensions}
\label{dfp}

The steady state of this system can be found exactly by noting that
the hopping rates Eq.(\ref{rates}) in this model correspond to a special
choice of rates in the Zero Range Process
\cite{spitzer,evansR}. In this process, a particle at site $k$
hops to its nearest neighbour independent of the state at the target
site, so that the interaction has zero range.
The steady state measure of this process can be found in any dimension
with or without bias. A convenient way to find it is by using the
condition of pairwise balance \cite{pwb} which states 
that for any given configuration $C$, one can find a configuration
$C^\prime$ in one-to-one correspondence with $C^{\prime \prime}$ such that
\be
W(C^\prime \rightarrow C) \; P(C^\prime)=W(C \rightarrow C^{\prime
\prime}) \; P(C)  \;\;\;\;,
\ee
where $P(C)$ is the probability of a configuration $C$ in a system with
$N$ sites and $W(C \rightarrow C^\prime)$ is the transition rate from
$C$ to $C^\prime$. The above condition is satisfied by
\be
P(C)=\frac{1}{\cal{N}} \prod_{k=1}^{N} f_{k}(m)  \;\;\;\;,
\ee
where
\bea
f_{k}(m)&=&\prod_{i=1}^{m} \frac{1}{u_{k}(i)} \;\;\mbox{for} \;\;
m > 0 \no \\ 
      &=& 1 \;\; \mbox{for} \;\; m=0    \;\;\;\;,
\eea
and $u_k(m)$ is the rate at which a particle hops out of a site
$k$ having $m$ particles and $\cal{N}$ is the normalisation constant.
$f_k(m)$ is defined upto a multiplicative factor $v^m$
where $v$ can be interpreted as the fugacity.

For our model, using the above solution for the particle distribution
$f_k(m)$ and Eq.(\ref{rates}), we find that 
the normalised probability distribution $P_k(m)$ that the site $k$ in
the bulk has $m$ particles has the Poisson form,  
\be
P_k(m)=P(m)=\frac{e^{-v} v^m}{m!}     \;\;,\;k \neq 0  \l{bulkf}  \;\;\;\;.
\ee
The fugacity $v$ will be determined
using the conservation law. At the chipper site, one can obtain
solution for $P_0(m)$ when  $v < w$,  
\be
P_{0}^{NA}(m)= \left( 1-\frac{v}{w} \right) \left( \frac{v}{w} \right)^m
\;\;,\;\; \frac{v}{w} <1 \l{chipf}  \;\;\;\;,
\ee
where NA in the superscript stands for the Non-Aggregate phase which
we have added in the anticipation of a phase transition.

Using equations Eq.(\ref{bulkf}) and Eq.(\ref{chipf}), one can easily see
that the average particle number at site $k \neq 0$ is $\av{m_k}=v=w s_0$ 
which further leads to constraint equation Eq.(\ref{cons}). 
Since this constraint equation is identical to that in the aggregation
model, then as discussed in 
Section~\ref{dap}, the steady state of this model also has three
possible phases (see Fig.~\ref{s0}). But, as we will see, this
system never exists in the UA phase and there is a phase transition
from the NA phase to the PA phase as $\rho$ is increased, keeping $w$
fixed, in all dimensions, irrespective of bias. 

Solving for the average number of particles at the origin $\av{M_0}$
using Eq.(\ref{chipf}) and substituting in Eq.(\ref{cons}), we obtain 
\be
\frac{s_0}{1-s_0}= \rho L^d -w s_0 (L^d-1) \;\;,\;\; 0 \leq s_0 < 1
\;\;\;\;. \l{consfree}
\ee
To leading order in $L$, one obtains two
solutions, namely, $s_0=\rho /w$ and $s_0=1$. Since the above
analysis holds only when $s_0$ is strictly less than $1$, the only
valid solution is $s_0=\rho/w$ for $\rho < w$. For $\rho \geq w$,
$s_0$ is pinned to its 
maximum value $1$. Thus there is a phase transition from the NA phase
to the PA phase at $\rho_c=w$, as $\rho$ is increased, keeping $w$ fixed.

One can solve for $P_0(m)$
in the PA phase and at the critical point using the conservation law
Eq.(\ref{cons}) and the fact that Eq.(\ref{bulkf}) is valid for all $s_0$ in
the range [0,1]. Since the total number of particles is conserved, 
$P_0(m)= \sum_{C} P(C) \;\; \delta( \sum_{k \neq 0} m_{k}=M-m)$. 
The quantity on the RHS can be calculated strightforwardly. The result
is
\bea
P_0^{PA}(m)&=&\sqrt{\frac{1}{\pi \av{m^2} (L^d-1)}} \;\; \mbox{exp}
\left({-\frac{(M-m-\langle m \rangle (L^d-1))^2}{(L^d-1) \av{m^2}}}
\right) \;\;\;\;, \l{PAf}
\eea
where $\langle m \rangle$=$v$ and $\av{m^2}$=$v (1+v)$ from Eq.(\ref{bulkf}). 
One can see from the above equation that $\langle M_0
\rangle=M-w(L^d-1) \sim L^d$ 
and $P_0^{PA}(0)=1-s_0 \sim e^{-L^d}$ in the PA phase. 
Further, at the critical point, we have $\rho L^d = w (L^d-1)$ using
which in Eq.(\ref{PAf}), one obtains $P_0(m)$ at the critical point,
\be
P_0^{cp}(m)=\sqrt{\frac{4}{\pi \av{m^2} (L^d-1)}} \;\; \mbox{exp}
\left({-\frac{m^2}{(L^d-1) \av{m^2}}} \right)  \l{cpf} \;\;\;\;,
\ee 
which gives a power law decay in $L$ for $P_0^{cp}(0)$. 

The phase transition is brought about in this model in the following way.
In the NA phase which is a low density phase, the typical number of
particles at all sites including the 
chipper is of $O(1)$. As the density is increased, there is a phase
transition to the PA phase at $\rho_c=w$. In this high density phase,
each site in the bulk still supports only $O(1)$ number of particles
in accordance with Eq.(\ref{bulkf}) so that the extra 
particles condense on the chipper giving rise to $\av{M_0} \sim L^d$.
The mechanism of phase transition in this model is similar to that in
Bose-Einstein condensation as was pointed out in \cite{bec} in
a similar $1d$ model which shows a phase transition with $u_k(m)=w \;
\delta_{k,0}+(1-\delta_{k,0})$.

\section{Extensive disorder}
\label{extens}

In this Section, we describe a possible scenario in the more interesting
and physically relevant situation when there is an extensive number of
chipper sites. These are assumed to be placed randomly, with quenched
random chipping rates  
$w_k$ at site $k$ distributed according to a distribution
$\mbox{Prob}(w_k)$. We are interested primarily in the aggregation
model but as argued below, on large space and time scales the
behaviour of this model with extensive disorder resembles that of the
corresponding free particle model (a generalization of the model in 
Section~\ref{free} with an extensive number of chippers) which is
exactly solvable. 

 Let us consider the diffusion-aggregation process with no bias and
take the initial condition to have a random distribution of 
masses. Then the finite concentration $c$ of chipper sites brings in 
new length and time scales into the problem, namely the mean spacing 
$l_c \sim c^{1/d}$ between the chippers and the associated
diffusion time $t_c=l_c^2$. Evidently, $l_c$ and $t_c$ define respectively 
the relevant length and time scales over which the diffusing mass
clusters sense the presence of an extensive number of defects.

Let us consider a finite but low concentration of chipper sites so
that $1 \ll l_c \ll L$.
On time scales $t \ll t_c$, we would expect the system
to behave roughly as independent, finite systems of size $l_c$ with
typically a single chipper each. The typical state on these  short time scales
thus resembles the UA-like steady state discussed in Section~\ref{mft}.
The typical mass of clusters, both mobile and
localised is limited by the time and grows proportional to $t^{d/2}$
as long as $t \ll t_c$. 

As the time crosses $t_c$, the finite spacing between the chippers becomes
relevant. The primary effect is to limit the size of the aggregates
formed in the non-chipper region by the diffusion-aggregation process
to $\sim$ $l_c^d$, since on a time scale $\gtrsim t_c$ the mass
cluster is likely to encounter a chipper and get trapped. For $t \gg
t_c$, the coarse grained view of 
the system is that of mass exchanged between close-by random rate
chippers with each exchange taking a time $\sim$ $t_c$.
To the extent that only finite aggregates (with mass $\sim$ $l_c^d$
$\ll L^d$) 
are formed in the transit between chippers, it is plausible that on
large length and time scales, we may
ignore interaction effects (i.e. coalescence) and think of the system as
effectively free-particle like in the non-chipper region. 

The free particle model with extensive disorder is solvable along the same
lines as the single-chipper problem described in
Section~\ref{free}. It defines a Zero Range Process with the hopping rates,
\bea
u_k(m) &=& w_k \;\;,\;\; \mbox{if $k$ is a chipper site} \\ \no
	&=& m \;\;,\;\; \mbox{if $k$ is not a chipper site}
\eea
Let $x$ denote the fraction of chipper sites. We recover the
single-chipper model as $x \rightarrow 0$, while $x \rightarrow 1$
corresponds to every site being a chipper site, and is the model 
considered in \cite{kf,bec}. For all $x$, in the low-density phase, the mass 
distribution on the non-chipper sites follows the probability distribution
of Eq.(\ref{bulkf}), while at chipper sites, Eq.(\ref{chipf}) is valid 
with $w$ replaced by $w_k$.  Let $s_k$ be the occupation probability of
site $k$, and $s_0$ refer to the occupation probability of the site
with the lowest chipping rate, $w_0 = Min\{w_k\}$. Then in the steady
state, the spatial uniformity of the current leads to (i)
$w_ks_k=w_0s_0$ if $k$ is a chipper site,  
and  (ii) $\langle m_k \rangle = w_0 s_0$ if $k$ is not a chipper site.  Using 
these relations, one can write the mass conservation equation analogous to
Eq.(\ref{consfree}) as
\be
\frac{1}{L^d}\frac{s_0}{1-s_0}+x \int{ dw \frac{\mbox{Prob}(w) s_0}{w/w_0
-s_0}}+(1-x) s_0 w_0=\rho  \;\;\;\;.  \l{exten}
\ee
where we have separated out the first term corresponding to the
slowest chipper. 

One can analyse Eq.(\ref{exten}) for various $x$ as follows:
(a) In the limit of a single chipper (reached as $x$ approaches $0$), we know
from Section~\ref{free} that as the density is increased, there is a 
phase transition from the NA phase to the PA phase with an infinite 
aggregate at the chipper site. 
(b) In the limit $x \rightarrow 1$, the 
model reduces to that considered in \cite{kf,bec}, where it is shown
that for Prob$(w_k) \sim (w_k-w_0)^{\nu}$
as $w_k \rightarrow w_0$, the system stays in the NA
phase for all $\rho$ if $\nu \leq 0$. On the other hand, if
$\nu > 0$, then there is a transition 
to the PA phase with an infinite aggregate at the site with chipping 
rate $w_0$ when the density crosses the critical density given by
\be
\rho_c(x=1) = \int {dw \frac{\mbox{Prob}(w)}{w/w_0 -1}} \;\;\;\;.
\ee
(c) For $0 < x < 1$, there is a transition from the NA to the PA phase
if Prob$(w)$ is chosen as in (b) above. The critical density can be 
determined by  taking
 $s_0 \rightarrow 1$ and $L^d (1-s_0) \rightarrow \infty$ in Eq.(\ref{exten}),
with the result
\be
 \rho_c(x) = x \rho_c(x=1) + (1-x) w_0  \;\;\;\;.
\ee
Thus the critical density interpolates linearly between its values in
the limits $x=0$ and $x=1$.

In view of the correspondence discussed at the beginning of this Section, we
would expect the aggregation model with a similar distribution of
chipper sites to show a phase transition from the NA to the PA phase.
In the former phase, there are both localized and mobile aggregates
with typical mass $\sim l_c^d$. In the PA phase, the distribution of masses 
is similar to that in the NA phase at all sites except the slowest
chipper with $w_k = w_0$; at this slowest site, there is an aggregate
with mass of order volume.

\section{Summary}
\label{concl}

In this paper, we introduced a minimal model to study the effect of
quenched, sitewise disorder in a aggregation-fragmentation system. Our
model had some simplifications: the fragmentation was allowed to
occur only at the trapping sites, and mass-independent
kernels for aggregation and fragmentation were considered. Despite
these simplifications, it retains the important physical
effects of diffusion, aggregation, fragmentation and trapping.

We studied the case of a single-chipper aggregation model in
detail. In all cases except the $1d$, biased case, 
the system exists in the UA phase in
which an infinite aggregate localised at the chipper site can coexist
with mobile infinite aggregates in the bulk. The simultaneous
existence of more than one infinite aggregate is a new feature absent in 
previous studies of translationally invariant systems. In the $1d$,
biased case, there is a phase transition
from a phase in which a localised aggregate is formed at the
chipper site (PA phase) to the one in which no aggregate is formed
anywhere in the system (NA phase) as the density is decreased. 

We also studied a variant of the above aggregation model in which
particles chip off at a single site but diffuse freely in the
bulk. This model can be solved exactly and shows a phase transition
from the PA phase to the NA 
phase as density is increased in all dimensions for all bias.

Finally we discussed a likely scenario for the aggregation model in
the presence of extensive disorder and argued that interaction
effects arising due to coalescence can be ignored on large enough time
scales. It would be interesting 
to check this expectation by a more detailed study of the system
with extensive disorder.

{Acknowledgements:}
We thank D. Dhar and Rajesh R. for useful discussions and comments on
an earlier version of the manuscript.



\begin{thebibliography}{999}
\bibitem{comment}
For the $1d$, biased case, the MFT result for mean-squared mass is
$\av{m_k^2}=Lf(k/L)$ where $f(x) \sim x$. Thus MFT is
sensitive to bias in this case unlike the translationally invariant cases (see
\cite{cmam} for example) in which the mean field equations are
independent of the bias.
\end{thebibliography}
\end{document}